\documentclass[twocolumn,amsmath,amssymb,aps]{revtex4-1}
\makeatletter
\newcommand*{\rom}[1]{\expandafter\@slowromancap\romannumeral #1@}
\makeatother
\usepackage{graphicx}
\usepackage{dcolumn}
\usepackage{bm}
\usepackage[noend]{algpseudocode}
\usepackage[ruled,vlined]{algorithm2e}
\usepackage{amsmath,xparse,mleftright}
\usepackage{subfigure}
\usepackage{braket,color}

\begin{document}

\title{Fibonacci anyons versus Majorana fermions}%

\author{Emil G\'enetay Johansen}
\author{Tapio Simula}
\affiliation{Optical Sciences Centre, Swinburne University of Technology, Melbourne 3122, Australia\\
}

\begin{abstract}
We have studied ${\rm SU}(2)_k$ anyon models, assessing their prospects for topological quantum computation. In particular, we have compared the Ising ($k=2$) anyon and Fibonacci ($k=3$) anyon models, motivated by their potential for future realizations based on Majorana fermion quasiparticles or exotic fractional quantum-Hall states, respectively. The quantum computational performance of the different anyon models is quantified at single qubit level by the difference between a target unitary operator and its approximation realised by anyon braiding. To facilitate efficient comparisons, we have developed a Monte Carlo enhanced Solovay--Kitaev quantum compiler algorithm that finds optimal braid words in polynomial time from the exponentially large search tree. Since universal quantum computation cannot be achieved within the Ising anyon model by braiding alone, we have introduced an additional elementary phase gate to model a non-topological measurement process, which restores universality of the anyon model at the cost of breaking the full topological protection. We model conventional kinds of decoherence processes algorithmically by introducing a controllable noise term to all non-topological gate operations. We find that for reasonable levels of decoherence, even the hybrid Ising anyon model retains a significant topological advantage over a conventional, non-topological, quantum computer. Furthermore, we find that only surprisingly short anyon braids are ever required to be compiled due to the gate noise exceeding the intrinsic error of the braid words already for word lengths of the order of $100$ elementary braids. We conclude that the future for hybrid topological quantum computation remains promising.
\end{abstract}

\maketitle

\section{\label{sec:level1}Introduction}
Topological phases of matter have attracted a significant amount of attention in recent times due to the diversity of emerging physical phenomena heralded by them \cite{wen2017colloquium,witten2015three,kitaev2009topological,bombin2010topological,laughlin1983anomalous}. Systems with only two spatial dimensions have proven to be particularly rich as their excitations do not adhere to the spin-statistics theorem that conventionally divides all particles into bosons or fermions. This theorem is one of the great triumphs of (3+1)-dimensional quantum field theory, yet its validity breaks down in lower dimensional systems with far reaching consequences. The proof of this statement was put forward in 1977 by Myrheim and Leinaas \cite{leinaas1977theory}, propelling significant new activity in this field. Twenty years later, in 1997, Kitaev conceived the idea that such exotic phases of matter may hold the key to fault tolerant quantum computation \cite{kitaev2009topological}. 

Ever since Feynman proposed the concept of quantum computation  \cite{feynman1982simulating}, making it a reality has been an aspiration for many. As with most new technologies, many hurdles were soon encountered, some of which are yet to be resolved. Despite of this, the first prototype quantum computers have recently been introduced. The very essence of quantum computation is to exploit the superposition principle and entanglement in quantum systems and encode information in the vast spaces that quantum states inhabit. By using quantum states as representations of information, it is possible to store and process exponentially larger quantities of information in comparison to what can be achieved using a conventional computer. 

However, quantum states are delicate and decohere upon interactions with the environment, which is why building such a computer is inherently challenging as the susceptibility to environmental noise will cause the information encoded to become distorted \cite{zurek2006decoherence,bergou2013introduction}. To remedy this, error correcting protocols that  can be utilised to restore the distorted strings of information to their original states were introduced \cite{kitaev1997quantum,knill2000theory,schumacher1996quantum}. Developing such protocols is indeed very challenging which motivated the search for systems that would be naturally immune to such forms of decoherence. In other words, systems which exhibit a type of symmetry that serves as a guard against external interactions were sought. For instance, a state of a ferromagnet is invariant under the $\rm{SU}(2)$ group operations but this symmetry protection breaks in the presence of an external magnetic field. By contrast, topological states of matter are robust and remain protected against continuous deformations, and can only be broken if the overall topology changes. Kitaev realized the untapped potential for the use of such systems in the context of computation, which then gave birth to the idea of topological quantum computation (TQC) \cite{kitaev2003fault}. Within the TQC paradigm, error correcting schemes are not necessary in principle as the states used to encode information possess an intrinsic, topological, protection from external noise sources. The whole enterprise is predicated on the very principle of topological equivalence, that is, the configuration is left invariant under diffeomorphisms, which entails that the information can be safely encoded and processed. The quasiparticle excitations in these systems are called \emph{anyons} \footnote{The term \emph{anyon} was coined by Frank Wilczek due to its fractional statistics. That is, instead of just being able to pick up $\pm 1$ upon permutation, an anyon can pick up \emph{any} phase \cite{wilczek1982quantum}.} \cite{leinaas1977theory,kitaev1997quantum,kitaev2003fault,nayak2008non,lerda2008anyons,wilczek1990fractional} as they are not restricted to be bosons or fermions and instead realise a fractional statistics. 

The rest of this work is organised as follows. For the sake of completeness, in Sec.~\ref{sec:level1} we provide a brief theoretical background to anyons and TQC. Section~\ref{sec:SU2k models} outlines the rudiments of the ${\rm SU}(2)_k$ anyon models. In Sec.~\ref{sec:quantum compiling} we introduce our Monte Carlo enhanced Solovay--Kitaev quantum compiler algoritm, which we use in Sec.~\ref{sec:results} for assessing the prospects of the Ising anyon and Fibonacci anyon models for TQC. Section~\ref{sec:conclusions} provides a summary and conclusions.

\section{\label{sec:level1}Elements of topological quantum computation}

The theoretical landscape of anyon theory is incredibly rich and substantial with modular tensor category theory \cite{panangaden2010categorical,blass2016quantum,blass2016quantum,panangaden2010categorical,kong2014anyon} and topological quantum field theory (TQFT) \cite{witten1988topological,atiyah1988topological,dunne1999aspects} being the main pillars. Here we are primarily concerned with applications in the context of quantum computation. The transformations induced when exchanging two anyons are not restricted to a factor $\pm 1$ as for bosons and fermions, but instead an arbitrary phase may be acquired. This special attribute, along with the intrinsic topological protection possessed by topological states, constitute the motivational foundation for the pursuit of TQC. 
\begin{figure}[!b]
    \centering
    \includegraphics[width=0.9\columnwidth]{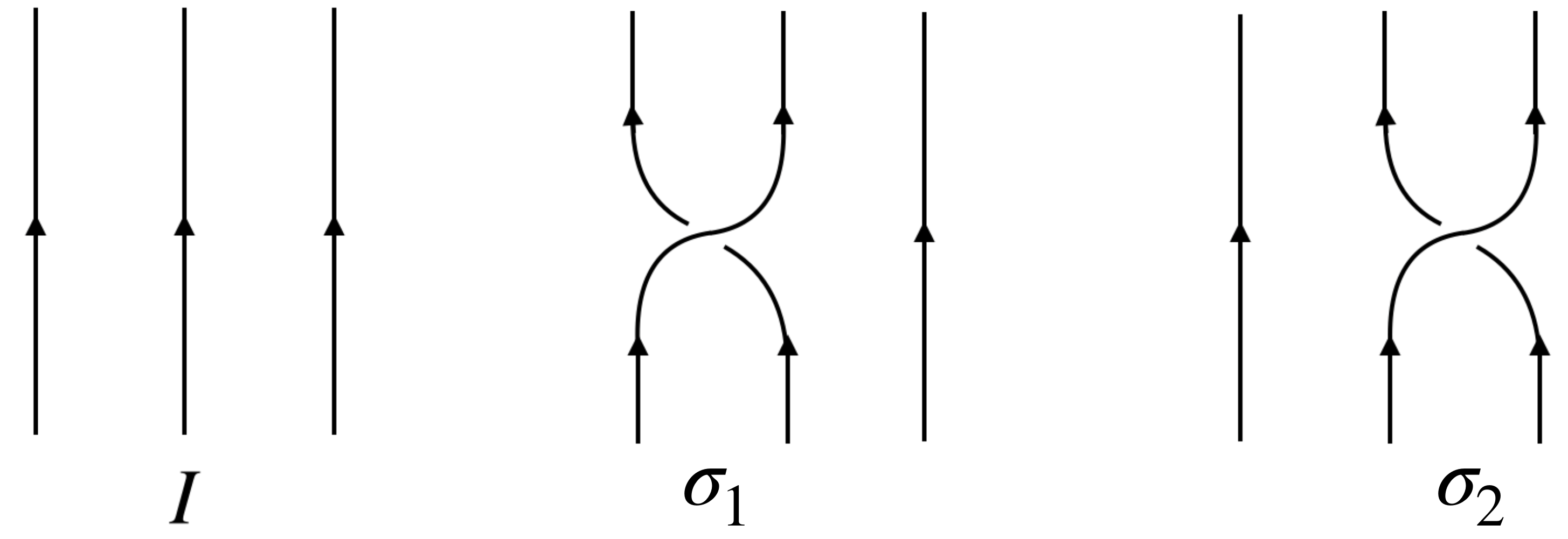}
    \caption{A pictorial representation of the generators of the braid group $\mathbb{B}_3$. The $\sigma_1$ operation swaps the positions of the first two anyons, $\sigma_2$ swaps the positions of the last two anyons and the identity operation $I$ does nothing to the system.}
    \label{generators}
\end{figure}

\begin{figure}[!b]
    \centering
    \includegraphics[width=0.6\columnwidth]{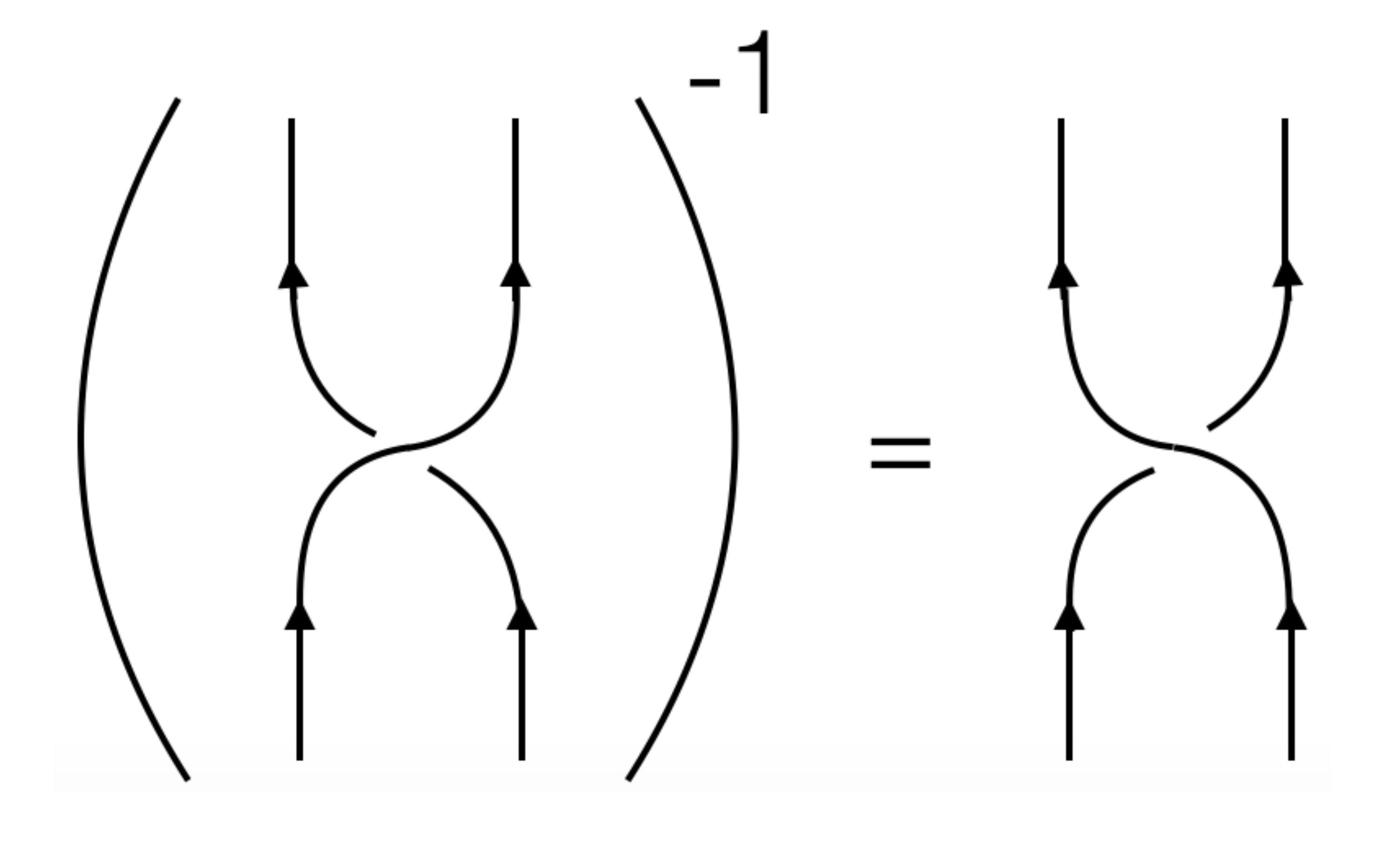}
    \caption{The inverse operation $\sigma^{-1}_i$ of the braid $\sigma_i$.}
    \label{inversebraid}
\end{figure}

When two anyons are exchanged the world lines will trace out a braid in $(2+1)-$dimensional space-time with two space and one time dimension. The set of braids together with the composition operation constitute a group known as the \emph{braid group} \cite{yang1994braid}. In particular, we will be working with the three stranded braid group $\mathbb{B}_3$ as this group can be used for encoding and processing 1-qubit quantum gates. Letting $\sigma_1$ denote the braid formed when the first anyon is wrapped around the second, and $\sigma_2$ denote the braid corresponding to the second wrapped around the third, the generator set $\Sigma_{\mathbb{B}_3}$ of $\mathbb{B}_3$ is $\Sigma_{\mathbb{B}_3} = \{ \sigma_1, \sigma_2, \sigma^{-1}_2, \sigma^{-1}_1 \}$. These generators are presented pictorially in Figs.~\ref{generators} and \ref{inversebraid} with the time flowing in the upward direction indicated by the arrows. These braid diagrams correspond to planar projections of the (2+1) dimensional anyon worldlines. When two lines intersect, the continuous line is understood to lay on top of the discontinuous one.

\begin{figure}[!t]
    \centering
    \includegraphics[width=0.6\columnwidth]{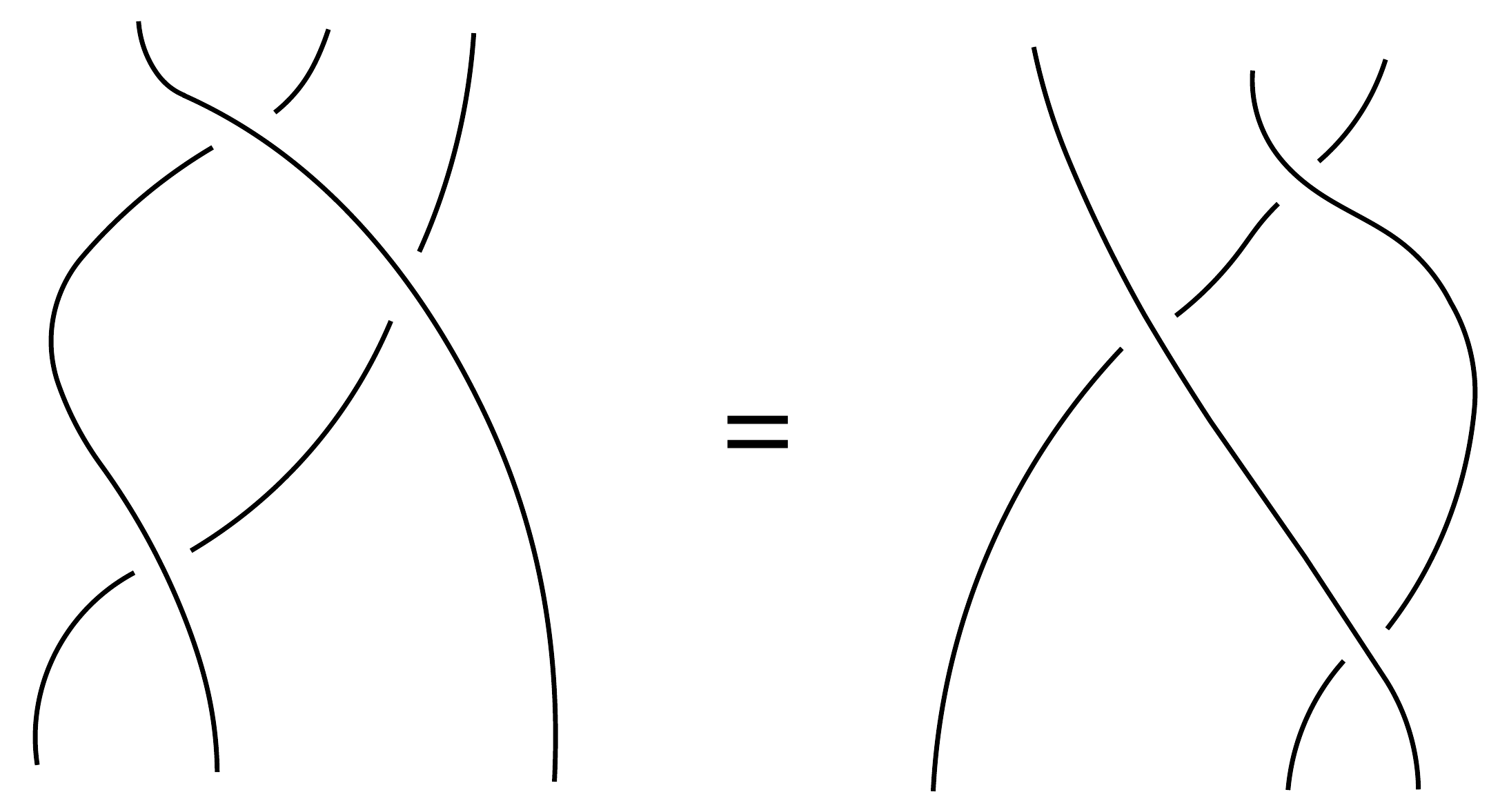}
    \caption{A graphical representation of the Yang-Baxter equation.}
    \label{YangBaxter}
\end{figure}

It is evident from these figures that when a braid is composed with its inverse by stacking the diagram on the left in Fig.~\ref{inversebraid} on top of the one on the right, or vice versa, the braid will ``untie" and become identical to the identity operation. If we consider the generic braid group $\mathbb{B}_N$ on $N$ strands with generators $\{ \sigma_i \}_{i=1}^{N-1}$, we can also conclude that any non-adjacent braids have to commute. In addition to this, the so called Yang-Baxter equation \cite{1989IJMPA...4.3759J} has to be satisfied for consistency. Algebraically, these conditions may be expressed as
\begin{equation}
    \centering
    \left\{
	\begin{array}{ll}
		\sigma_i \sigma_j = \sigma_j \sigma_i \,\  & \mbox{if } |i-j| \geq 2 \\
		
		\sigma_i \sigma_{i+1} \sigma_i = \sigma_{i+1} \sigma_{i} \sigma_{i+1}\,\ & \mbox{if } 1 \leq i \leq N-2\\
		
		\sigma_i \sigma_i^{-1} = \sigma_i^{-1} \sigma_i = I.
	\end{array}
    \right.
    \label{YBeqns}
\end{equation}
Figure~\ref{YangBaxter} shows a graphical representation of the Yang--Baxter equation. It is clear that when following the world lines, this condition has to be imposed in order to maintain consistency. The set of equations in Eq.~\eqref{YBeqns} admits many matrix solutions. The simplest one corresponds to the one dimensional representation and yields a trivial phase factor $e^{i \phi}$. However, there are also solutions constituting multi-dimensional representations, which give rise to a richer structure \cite{pachos2012introduction,wilczek1990fractional}.

Thus far we have only discussed anyons and the emergence of quasi-particles with fractional statistics in two dimensions generically. Anyons can further be categorized into abelian and non-abelian sub-species. These two sub-species are distinguished by the commutativity of the corresponding braid group they transform under; abelian anyons transform under an abelian braid group representation that is one dimensional, and non-abelian anyons under a higher dimensional non-abelian braid group representation. The dimensionality of the corresponding representation space has important consequences for what happens when multiple anyons are involved. If we consider two anyons and perform a dilation transformation, these quasi-particles may be regarded as one composite object. This process is known as \emph{fusion}. Due to the simple structure of abelian anyons, the fusion outcome is always definite, whereas for the non-abelian anyons it is indefinite. Considering the fusion product of two non-abelian anyons with charges $\alpha$ and $\beta$, they may, for instance, fuse into the vacuum denoted by $\mathbf{1}$ or a quasi-particle with charge $\psi$.  Mathematically, this is expressed as
\begin{equation}
    \alpha \otimes \beta = {\mathbf 1} \oplus \psi.
    \label{decomp}
\end{equation}
The decomposition rule in Eq.~\eqref{decomp} arises entirely due to the reducibility of the joint representation spaces in the non-abelian case, whereas the simple one dimensional structure of abelian anyons have no non-trivial invariant sub-spaces, thus resulting in a definite fusion outcome. 

In TQC, the core idea is to utilize the fusion product as a qubit in which information can be encoded. Moreover, since the state of the anyons can be transformed by braiding them, the corresponding braid group may be used for producing a circuit of quantum gates. However, it is not possible to use an arbitrary braid group representation. In order to produce any unitary target gate to arbitrary accuracy, the image of the representation must form a set that is topologically dense in ${\rm SU}(2)$. This implies that  all transformations in ${\rm SU}(2)$ are either contained within the braid group, or are limit ``points" to at least one element in the braid group. If this condition is satisfied the model is said to be \emph{computationally universal} \cite{dawson2005solovay}. Universality is crucial as it allows realization of any target logic gate in topologically protected way by braiding alone. A computational process is thus initiated by creating pairs of non-abelian anyons from the vacuum, followed by a particular series of braids that have been determined in advance based on the chosen quantum circuit. At the end of the braiding, the anyons are brought (sequentially fused) together, which is the read out. Taking the fusion product in Eq.~\eqref{decomp} as an example, we can either measure a vacuum, or a composite object $\psi$, each one with a certain probability which normalize to 1. 
\newline
Perhaps the two most celebrated examples of anyons are the Ising anyon and Fibonacci anyon models. While the Fibonacci model is universal, the Ising anyon model is not. Although anyons are elusive, it is believed that they might emerge as quasiparticle excitations in some specially engineered condensed matter systems. However, current research indicates that Ising anyons may be a far more accessible platform for TQC. As it turns out, the Majorana fermion is an incarnation of the Ising anyon, i.e., it obeys the same braiding and fusion rules. Majorana fermion quasiparticles are expected to be found in chiral $p$-wave paired Fermi superfluids \cite{Gurarie2007a,Tewari2007a,Mizushima2008a,Jiang2011a}, topological superconductors \cite{Volovik1999a,sato2016majorana, livanas2019alternative}, semi-conducting nano wires \cite{stanescu2013majorana} and certain quantum-Hall fluids \cite{kasahara2018majorana, zuo2016detecting} if we descend dimensions $d<3$. Fibonacci anyons, on the other hand might be out of reach for the time being, although it is thought that they could exist in some very exotic fractional quantum Hall fluids with filling fraction $\nu = 12/5$ \cite{mong2015fibonacci, xia2004electron}. 

One way to address the universality problem with the Ising anyon model is to add an additional phase gate to the set of braids. Such a phase gate does not have any intrinsic protection and is thus susceptible to noise. This means that the Ising anyon model is at best a hybrid between a fully protected topological quantum computer and a conventional one. In this paper we are particularly interested in evaluating this hybrid solution by comparing it to the topologically protected Fibonacci model. The two most pertinent questions we are addressing are how these two models perform compared to one another when a phase gate is added to the Ising generator set, and how severe the effect of noise is compared to a conventional quantum computer? 

The Ising anyon and the Fibonacci anyon models are special cases of a more general family of anyon models known as ${\rm SU}(2)_k$ models (${\rm SU}(2)$ at level $k$). Here the Ising anyon model corresponds to $k=2$ and the Fibonacci model to $k=3$. In addition to these, we are also including the $k=4,5,6,8$ models in our analysis. Like the Ising anyon model, the $k=4$ model is also non-universal, which is why we supplement it with a non-topological phase gate. The tool we employ to assess the performance of the anyon models is known as a \emph{quantum compiler}. A compiler is an algorithm that takes a target logic gate as an input and returns an approximation to this gate by building it from the generators of the particular anyon model. Provided a metric, the aim of a compiler is to construct a sequence of generators which results in a matrix that lies as close as possible to the target matrix. In our assessment we  first generate samples of random unitary matrices as target gates for which the compiler then constructs approximations. We then compare the results for the various anyon models to understand how well they are performing with respect to each other.

\section{${\rm SU}(2)_k$ anyon models}
\label{sec:SU2k models}
In the pursuit of quantum computation a wide range of mathematical models have been suggested. Some of the most notable anyon models are the so called ${\rm SU}(2)_k$ models. These models correspond to various truncations of the spin chain expansion that will be discussed in Sec.~\ref{subsec:SU2k}. The underpinning mathematical framework describing the ${\rm SU}(2)_k$ models is called topological quantum field theory(TQFT) (in particular Chern--Simons theory), and it provides us with tools to study physical processes in topological phases of matter. The Aharonov--Bohm experiment \cite{berry1984quantal,aharonov1959significance} is one of the clearest realisations of such topological field theories. This experiment confirms the mechanism that gives rise to an abelian phase, due to the underlying abelian field theory, and the notion of a vector potential can be generalised to arbitrary two dimensional systems as an emergent phenomenon due to the punctured manifold the system is defined on. Moreover, the vector potential may be promoted to a tensor of higher rank, which in turn gives rise to a non-abelian theory. 

Non-abelian ${\rm SU}(2)_k$ theories are gauge invariant only up to a phase $2\pi n k$, where $n$ is the winding number. When computing the amplitude for a given process, this quantity is gauge invariant only if $k$ is an integer. Hence, $k \in \mathbb{Z}$ is labelling the theory, and is referred to as the level of the theory. For more detailed descriptions of TQFTs and ${\rm SU}(2)_k $ theories in particular, see e.g \cite{witten1988topological,fradkin1998chern,gawelenedzki19912}.

\subsection{Deformed ${\rm SU}(2)$ spin algebras}
\label{subsec:SU2k}
In quantum mechanics the  joint tensored space
of interacting spins
can be decomposed into a direct sum of orthogonal irreducible sub-spaces. In particular, for spins $\textbf{S}_1$ and $\textbf{S}_2$ this is expressed as
\begin{equation}
    \textbf{S}_1 \otimes \textbf{S}_2 = |\textbf{S}_1 - \textbf{S}_2| \oplus \dotsb \oplus (\textbf{S}_1 + \textbf{S}_2).
\end{equation}
If we consider all representations of ${\rm SU}(2)$, the spins may take any integer or half integer value, i.e. $S_i = \frac{1}{2},1,\frac{3}{2},2,..$. This sequence continues to infinity, but what if we terminate it after a specific value? This is essentially what is meant by deforming the algebra as only a subset of all representations are allowed. Thus, if we consider generalized angular momenta $j_i$ and choose some truncation level $k$, the corresponding algebra decomposes as \cite{gils2013anyonic}
\begin{align}
    j_1 \otimes j_2 &=
    |j_1 - j_2| \oplus (|j_1 - j_2|+1)\oplus \dotsb\notag\\
    \dotsb &\oplus {\rm min}(j_1 + j_2, k - j_1 - j_2),
    \label{frule}
\end{align}
with $l = \frac{1}{2},1,\frac{3}{2},2,...,\frac{k}{2}$ the allowed values for the generalised spin, and $k \to \infty$ corresponding to the full ${\rm SU}(2)$ algebra. 

For the sake of concreteness, let us consider two spin-$\frac{1}{2}$ particles for which
\begin{equation}
    \frac{1}{2} \otimes \frac{1}{2} = 0 \oplus 1.
\end{equation}
In words, two spin-$\frac{1}{2}$ spaces, decompose as a spin-$0$ (singlet) space and a spin-$1$ (triplet) representation space. This is the only non-trivial fusion rule for $k = 2$. Similarly, for two spin-$1$ particles at level $k = 3$, we have
\begin{equation}
    1 \otimes 1 = 0 \oplus 1
    \label{fibrule}
\end{equation}
since the spin-2 representation is now excluded. Note that the notion of spin and angular momentum is used here in a general sense. That is, it could be any type of `charge' obeying the same algebra. In this particular context, these labels represent topological charges of anyons and the tensor product represents the fusion of two anyons, i.e. their combined global charge. However, when $k$ is odd, there exists a fusion automophism, which defines a duality between the integer and half integer charges. In general we have the following map \cite{trebst2008short}
\begin{equation}
    \frac{k}{2} \otimes j = j',
    \label{autom}
\end{equation}
where $j$ has half-integer spin (or integer) and $j'$ has integer spin (or half-interger if $j$ has integer spin). 

\begin{figure}[!b]
    \centering
    \includegraphics[width=\columnwidth]{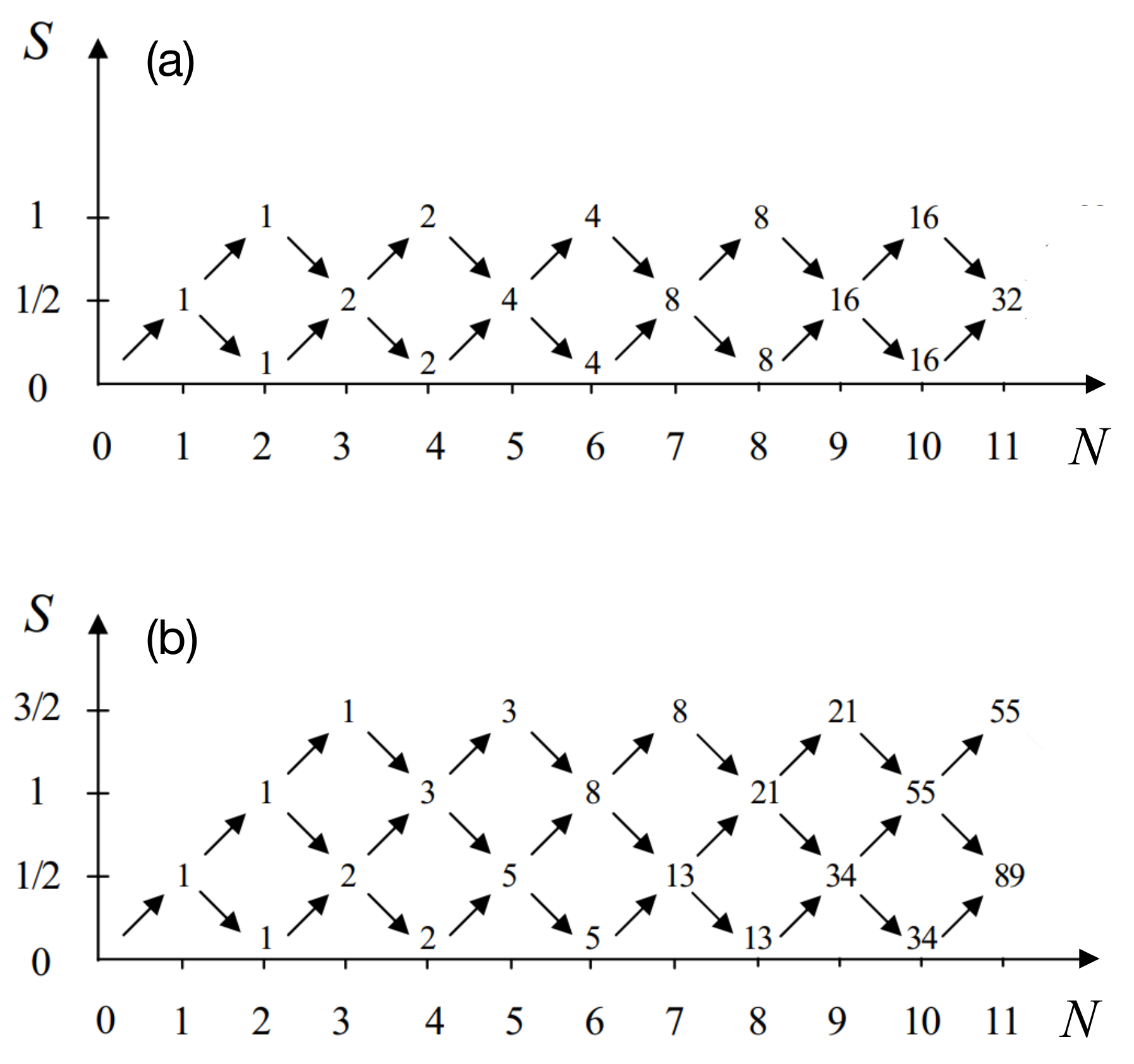}
    \caption{Two Bratteli diagrams for (a) $k=2$ and (b) $k=3$. The vertex numbers correspond to the degeneracy of the fusion outcome and thus the dimension of the Hilbert space. These numbers are determined by the number of paths from the origin leading to a particular vertex
    \label{bratteli}
    \cite{hormozi2007topological}.}
\end{figure}

In TQC we let the fusion product constitute a qubit. First, anyon pairs are created from the vacuum, then the computation is carried out by braiding them, and finally they are brought together (fused), which corresponds to the read out. For instance, in Eq.~\eqref{fibrule}  one would measure a spin equal to 0 with a probability $p_0$, or 1 with probability $p_1$, where $p_0 + p_1 =1$. The degeneracy of the fusion outcome for an arbitrary $N$-anyon system can be neatly summarized using a Bratteli diagram, see Fig.~\ref{bratteli} \cite{hormozi2007topological}.
Bratteli diagrams are very useful as they provide a simple way of determining the dimension of the fusion space. The dimension depends on the topological charge of the anyons, and is determined by the number of paths leading to that particular vertex. In general, the dimension of the fusion space of $N$ anyons can be computed as \cite{hormozi2007topological}
\begin{equation}
    \lfloor n \rfloor_q ^N= \left(\frac{q^{\frac{n}{2}} - q^{-\frac{n}{2}}}{q^{\frac{1}{2}} - q^{-\frac{1}{2}}}\right)^N,
    \label{qnum}
\end{equation}
where $n \in \frac{1}{2} \mathbb{Z}$ and $q =e^{i 2 \pi / (k+2)}$. This is referred to as a \emph{q-deformed} integer and it can be shown that all ${\rm SU}(2)$ representations are recovered, i.e. $\lfloor n \rfloor_q \longrightarrow n$  and thus $q \to 1$, as we let $k \to \infty$ \cite{delaney2016local}. In fact, this is the very reason why the Hilbert spaces spanned by multiple full ${\rm SU}(2)$ spins can be decomposed into irreducible sub-spaces since these systems correspond to $k \to\infty$, which yields an integer dimension. Thus, we can think of $\lfloor n \rfloor_q$ as a deformation of the integer $n$. When $k$ takes on a finite value, however, the deformed integer will in many cases be irrational, which means that such a decomposition is not always possible as we eliminated higher dimensional representation spaces, leaving a deformed structure behind. This non-trivial decomposition signals an existence of non-abelian braiding statistics, which will be discussed in more detail in Sec.~\ref{subsec:FandR}. 

\begin{figure}[!t]
    \centering
    \includegraphics[width=\columnwidth]{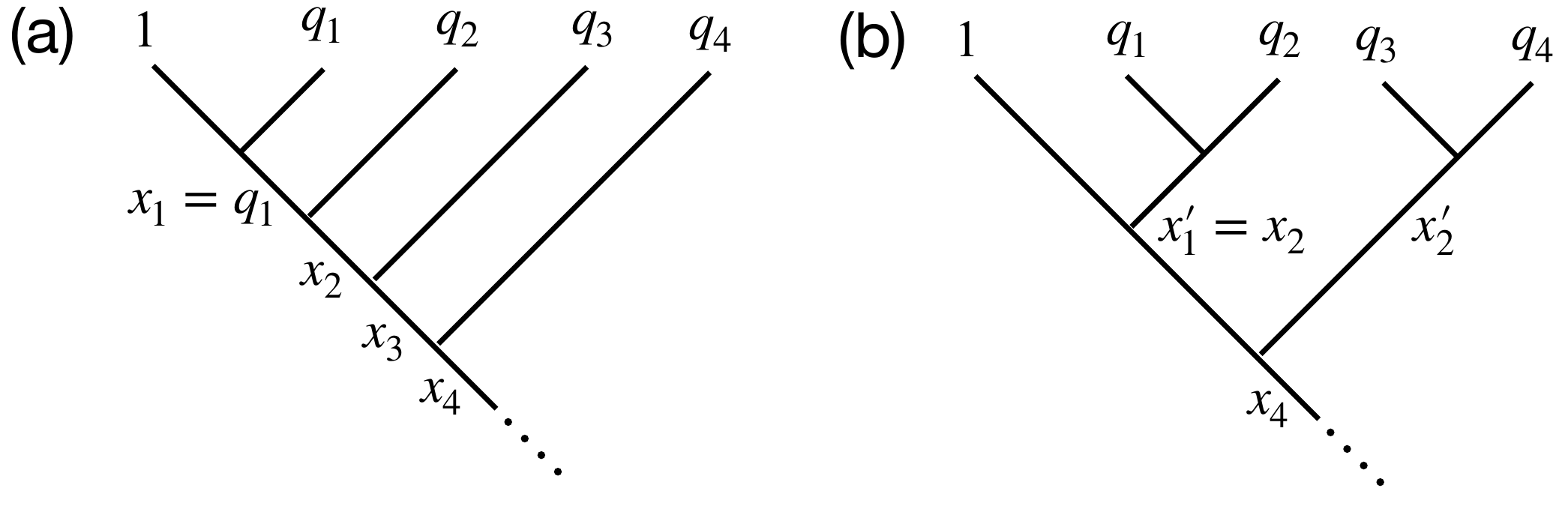}
    \caption{A fusion tree representation of the process $q_1 \otimes q_2 \otimes q_3 \otimes q_4$ where $1$ represents the vacuum.}
    \label{fusiontree}
\end{figure}

It is convenient to introduce a schematic notation for the fusion processes. Let us denote by $q_i$ the charge of the $i$'th anyon and by $x_i$ the fusion outcome of anyon $i$ and $i+1$, then a fusion process can be represented as a tree diagram. In Fig.~\ref{fusiontree}(a) a fusion process involving the charges $q_1$, $q_2$, $q_3$ and $q_4$ is presented. Henceforth we shall adopt this notation for fusion processes.

\subsection{F-moves and R-moves \label{subsec:FandR}}
\begin{figure}[!b]
    \centering
    \includegraphics[width=\columnwidth]{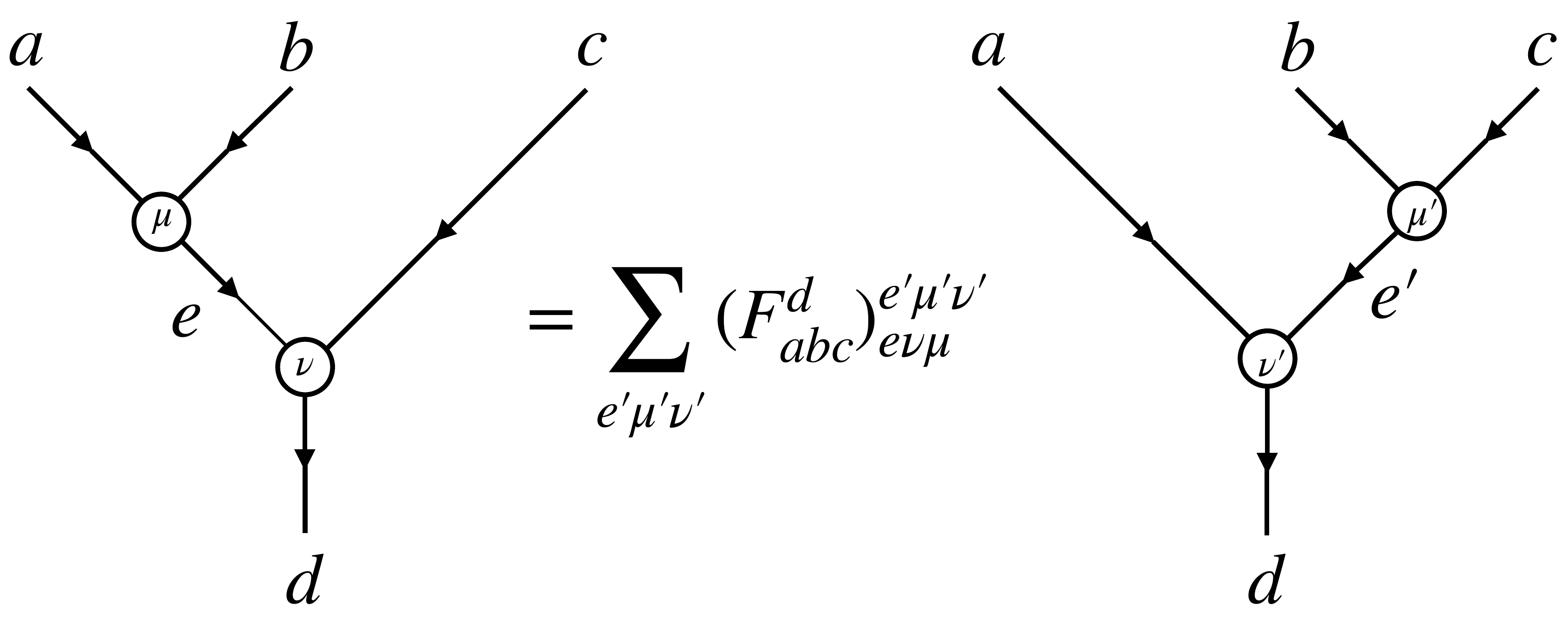}
    \caption{A graphical illustration of the F-move corresponding to the Eq.~(\ref{Fmoveq}).}
    \label{fmove}
\end{figure}
When anyons participate in a fusion process, the final outcome must be independent of the order in which the anyons are being fused, since the total charge must be preserved. Or in mathematical terms, we have to enforce associativity upon the fusion rules. For instance, in Fig.~\ref{fusiontree}(a), the anyons are fused sequentially, but we could just as well have fused $q_1$ with $q_2$ right away to form $x_1'$ (this is essentially what is happening already as $q_1$ is first fused with the vacuum, which implies that $x_1' = x_2$) and then $x_1' = x_2$ with the fusion outcome $x_2'$ of $q_3$ and $q_4$. This process is depicted in Fig.~\ref{fusiontree}(b).

In topological quantum computation the two configurations, Figs~\ref{fusiontree}(a) and (b), correspond to different basis states and a change from one basis to another is realised by applying an $F$-move. For simplicity, let us consider three anyons $a, b, c$ and their fusion product $d$, which can be formed in $N_{abc}^d$ distinct ways. Their fusion space $V_{abc}^d$ can then be decomposed in two different, but up to an isomorphism, identical ways such that \cite{preskill2004topological}
 \begin{equation}
    V^d_{abc} \simeq \bigoplus_e V^e_{ab} \otimes V^d_{ec} \simeq \bigoplus_{e'} V^{e'}_{bc} \otimes V^d_{ae'}.
 \end{equation}
First the tensor product of the space corresponding to $a$ and $b$ forming $e$, and $e$ and $c$ forming d, and second,  $b$ and $c$ forming $e'$ and $a$ and $e'$ forming $d$, where the outcomes $e$ and $e'$ are being summed over. We may also assign ket vectors to the different bases
  \begin{equation}
     \ket{(ab)c \rightarrow d;e, \mu, \nu } \equiv \ket{ab;e, \mu} \otimes \ket{ec;d, \nu}
  \end{equation} 
 and 
 \begin{equation}
     \ket{a(bc) \rightarrow d;e', \mu', \nu'} \equiv \ket{ae';d, \nu'} \otimes \ket{bc;e', \mu'}.
  \end{equation}
Here $\mu,\nu,\mu'$ and $\nu'$ label the specific fusion channels out of the $N_{abc}^d$ unique products. The transformation $F$ maps the basis $a \otimes (b \otimes c)$ to $(a \otimes b)\otimes c$ and can be formulated as
 \begin{equation}
    \ket{(ab)c \rightarrow d;e, \mu, \nu } = \sum_{e',\mu',\nu'}(F^d_{abc})^{e'\mu'\nu'}_{e\mu\nu} \ket{a(bc) \rightarrow d;e', \mu', \nu' },
    \label{Fmoveq}
 \end{equation}
or expressed graphically as shown in Fig.~\ref{fmove}.
Moreover, braiding two of the charges should not change the total charge. 
\begin{figure}[!b]
    \centering
    \includegraphics[width=0.75\columnwidth]{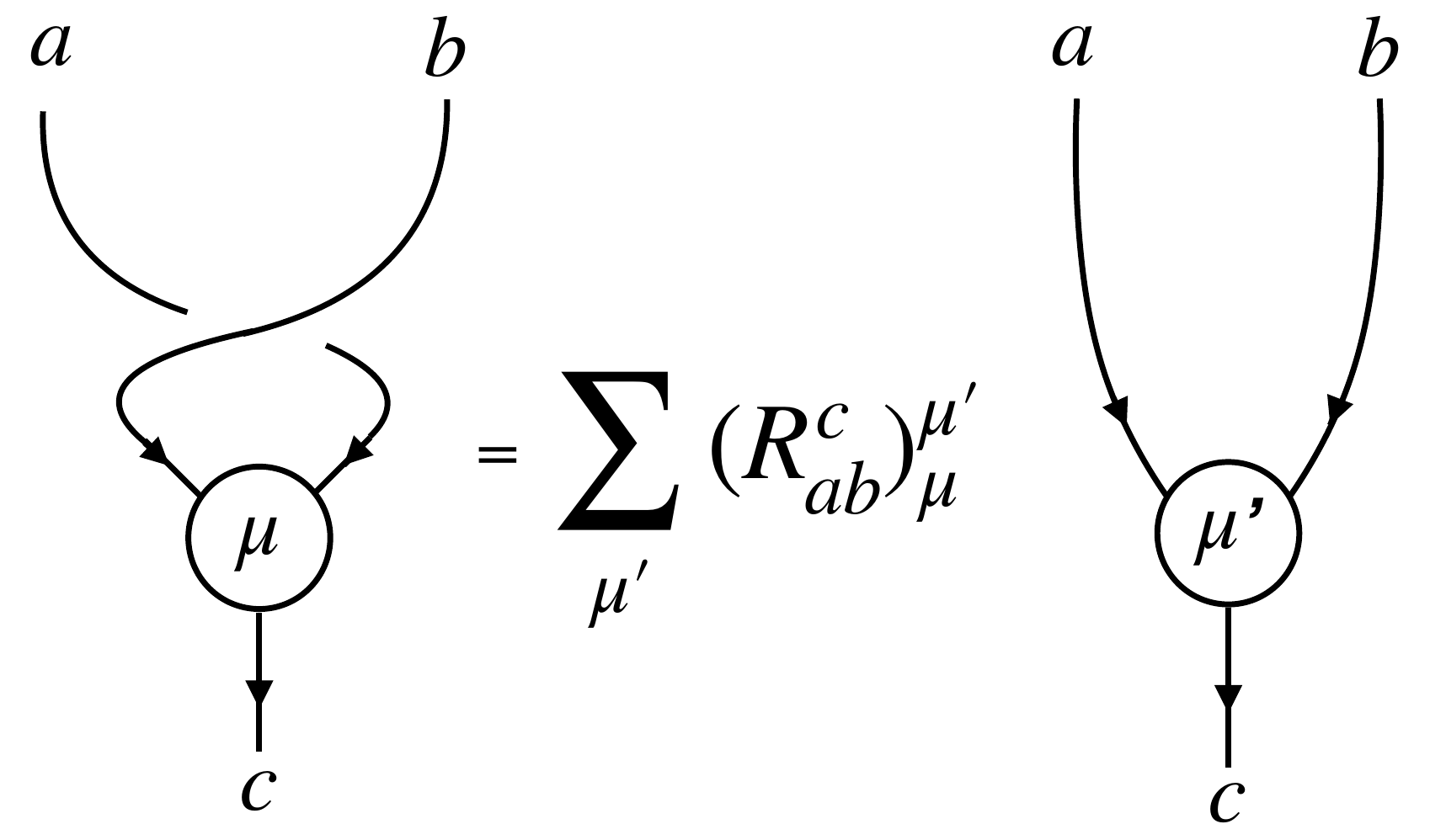}
    \caption{A graphical illustration of the R-move.}
    \label{rmove}
\end{figure}
Figure \ref{rmove} illustrates this charge conservation in terms of an $R$-move. By applying these transformations sequentially it is possible to derive two consistency equations known as the hexagon equation
\begin{align}\label{penta}
     &\sum_{\lambda\gamma}[R^{ac}_e]_{\alpha\lambda}[F^{acb}_d]_{(e\lambda\beta)(g\gamma\nu)}[R^{bc}_g]_{\gamma\mu}= \\
     &\sum_{f\sigma\delta\psi}[F^{cab}_d]_{(e\alpha\beta)(f\delta\sigma)}[R^{fc}_d]_{\sigma\psi}[F^{abc}_d]_{(f\delta\psi)(g\mu\nu)}\nonumber 
\end{align}
and the pentagon equation
\begin{align} \label{hexa}
     &\sum_{\delta}[F^{fcd}_e]_{(g\beta\gamma)(l\delta\nu)}[F^{abl}_e]_{(f\alpha\delta)(k\lambda\mu)}=\\
    &\sum_{h\sigma\psi\rho}[F^{abc}_g]_{(f\alpha\beta)(h\sigma\psi)}[F^{ahd}_e]_{(g\sigma\gamma)(k\lambda\rho)}[F^{bcd}_k]_{(h\psi\rho)(l\mu\nu)}\nonumber,
\end{align}
and a graphical representation of these equations can be found for instance in Refs.~\cite{nayak2008non,trebst2008short,field2018introduction}.

\begin{table}[!t]
\caption{Fusion table for $k=3$ anyon model.}
\begin{ruledtabular}
\begin{tabular}{ccc}
\textrm{$\otimes$}& $j_1$=0& $j_1=1$\\
$j_2=0$ & $0$ & $1$\\
$j_2 =1$ & $1$ & $0 \oplus 1$
\end{tabular}
\label{Fibotable}
\end{ruledtabular}
\end{table}
\begin{table}[!b]
\caption{Fusion table for $k=2$ anyon model.}
\begin{ruledtabular}
\begin{tabular}{cccc}
\textrm{$\otimes$}& $j_1$=0& $j_1=\frac{1}{2}$& $j_1=1$\\
$j_2=0$ & $0$ & $\frac{1}{2}$ & $1$\\
$j_2 =\frac{1}{2}$ & $\frac{1}{2}$ & $0 \oplus 1$ & $\frac{1}{2}$\\
$j_2 =1$ & $1$ & $\frac{1}{2}$ & $0$
\end{tabular}
\label{isingtable}
\end{ruledtabular}
\end{table}

The solutions to this set of equations embody all essential information about the anyon model since we are mostly interested in the tranformation properties under F-moves and R-moves. Suppose we would like to exchange charges $a$ and $b$ and that our initial basis is $\ket{a(bc) \to d}$. We would first have to transform the basis to $\ket{(ab)c \to d}$ under the action of $F$, in order to execute the exchange, and thereafter return to the initial basis by acting with the $F^{-1}$. Hence, in this new basis, the braid generators would be given by 
\begin{equation}
   B_{ab} = F^{-1} R_{ab} F   
   \label{bg1}
\end{equation}
and
\begin{equation}
    B_{bc} = R_{bc}.
     \label{bg2}
\end{equation}
The solutions to the hexagon and pentagon equations for an arbitrary value of $k$ have been worked out in comprehensive literature on quantum groups, see for instance Refs.~\cite{fuchs1995affine, kirillov1991representations, delaney2016local}.  These solutions are 
\begin{equation}
  R^{c}_{ab} =  (-1)^{(a+b-c)/2} q^{-[a(a+2)+b(b+2)-c(c+2)]/2}
  \label{req}
\end{equation}
and
 \begin{align}
     (F^{abc}_d)_f^e = (-1)^{a+b+c+d} \Delta (a,b,e) \Delta (c,d,e) \nonumber  \\
     \times \Delta (b,c,f) \Delta (a,d,f) \sqrt{\lfloor 2e + 1 \rfloor} \sqrt{\lfloor 2f + 1 \rfloor}\nonumber\\
     \times \begin{Bmatrix} 
      a & b & e \\ 
      c & d & f 
   \end{Bmatrix}
   \label{feq}
 \end{align}
where
\begin{align}\label{sixj}
     &   \begin{Bmatrix} 
      a & b & e \\ 
      c & d & f
   \end{Bmatrix}= 
    \sum_n \frac{(-1)^n \lfloor n+1 \rfloor}{\lfloor a+b+c+d-n \rfloor !} \times \\ 
   & \frac{1}{\lfloor a+c+e+f-n \rfloor ! \lfloor b+d+e+f-n \rfloor !\lfloor n-a-b-e \rfloor !}\nonumber\\
    & \times\frac{1}{ \lfloor n-c-d-e \rfloor ! \lfloor n-b-c-f \rfloor ! \lfloor n-a-d-f \rfloor !}\nonumber .
\end{align}
Equation~(\ref{sixj}) is a q-deformed version of the Wigner 6j-symbol describing the transformation under recoupling in the representation theory of ${\rm SU}(2)_k$.

\begin{table}[!t]
\caption{Fusion table for $k=4$ anyon model.}
\begin{ruledtabular}
\label{k4table}
\vspace*{2mm}
\begin{tabular}{cccccc}
\textrm{$\otimes$}& $j_1$=0& $j_1=\frac{1}{2}$& $j_1=1$ & $j_1=\frac{3}{2}$ & $j_1=2$\\
$j_2=0$ & $0$ & $\frac{1}{2}$ & $1$ & $\frac{3}{2}$ & $2$\\
$j_2 =\frac{1}{2}$ & $\frac{1}{2}$ & $0 \oplus 1$ & $\frac{1}{2}$ & $1 \oplus 2$ & $\frac{3}{2}$\\
$j_2 =1$ & $1$ & $\frac{1}{2} \oplus \frac{3}{2}$ & $0 \oplus 1 \oplus 2$ & $\frac{1}{2} \oplus \frac{3}{2}$ & $1$\\
$j_2 =\frac{3}{2}$ & $\frac{3}{2}$ & $1 \oplus 2$ & $\frac{1}{2} \oplus \frac{3}{2}$ & $0 \oplus 1$ & $\frac{1}{2} \oplus \frac{3}{2}$\\
$j_2 =2$ & $2$ & $\frac{3}{2}$ & $1$ & $\frac{1}{2} \oplus \frac{3}{2}$ & 0
\end{tabular}
\vspace*{2mm}
\end{ruledtabular}
\end{table}
\begin{table}[!b]
\caption{Fusion table for $k=5$ anyon model.}
\begin{ruledtabular}
\vspace*{2mm}
\begin{tabular}{cccc}
\textrm{$\otimes$}& $j_1$=0& $j_1=1$& $j_1=2$\\
$j_2=0$ & $0$ & $1$ & $2$\\
$j_2 =1$ & $1$ & $0 \oplus 1 \oplus 2$ & $1$\\
$j_2 =2$ & $2$ & $1$ & $0 \oplus 1$
\end{tabular}
\label{k5table}
\vspace*{2mm}
\end{ruledtabular}
\end{table}

\subsection{Universal quantum computation}

As previously mentioned, computational universality is required in order to carry out general purpose quantum computation. Computational universality is determined by whether the braid group under consideration is able to approximate the group constituted by all possible rotations of ${\rm SU}(2)$ on the Bloch sphere. The Solovay--Kitaev theorem, discussed in more detail in Sec.~\rom{4}.C, states that universality is achieved only if the gate set is generating a topologically dense cover in ${\rm SU}(2)$. Thus, an important question is for what values of $k$ is this satisfied? All information about the relation between the level $k$ and universality can be summarized in the following two theorems \cite{nayak2008non}:
\newline
\newline
\textbf{Theorem \rom{2.1}.} \textit{For $k \in \{ 1,2,4 \}$ $\rho_{k,n}(\mathbb{B}_n)$ is a finite group. For other values of $k$ and $n \geq 3$, $\rho_{k,n}(\mathbb{B}_n)$ is infinite, except for when k = 8 and n = 4}.
\newline
\newline
Here $\rho_{k,n}$ is a representation of the braid group $\mathbb{B}_n$ on $n$ strands. It was later shown by Freedman, Larsen, and Wang (2002) that the whole special unitary group will be contained in the closed image of this representation for these particular values of $k$, which led to the second fundamental theorem \cite{freedman2002modular}
\newline
\newline
\textbf{Theorem \rom{2.2}.} \textit{When the image of $\overline{\rho_{k,n}}$ is infinite it holds that {\rm SU}$(V_{k,n}) \subset image(\overline{\rho_{k,n}})$.}
\newline
\newline
In words, this simply entails that when the closed image is infinite, the special unitary group will be contained in the image, which further implies that universality is achieved. This is a very natural result as the Bloch sphere is a compact manifold. Hence, if the image contains an infinite number of unique points, the set must fill up the entire sphere densely.

\subsection{The Fibonacci and Ising anyon models}

The Fibonacci ($k=3$) and Ising  ($k=2$) anyon models are the simplest and by far the most studied ${\rm SU}(2)_k$ models. Physical realizations of Ising anyons have already been proposed in the form of Majorana fermion zero modes\footnote{The Majorana fermion was first predicted by the Italian physicist Ettore Majorana. Contrary to the Dirac fermions, these particles have the property of being their own anti-particles, which means that they will annihilate when two such particles are fused.}. Experimental evidence for Fibonacci anyons, on the contrary, have not been reported yet although they are believed to exist as quasiparticles in Kondo systems \cite{komijani2019kondo} and in exotic quantum Hall fluids with filling fraction $\nu = 12/5$ \cite{mong2015fibonacci, xia2004electron}. By virtue of the automorphism defined in Eq.~\eqref{autom}, the ${\rm SU}(2)_3$ Fibonacci anyon model only have integer (or half integer) charges. In particular, for $k=3$, we have $j = 0, 1$ which yields the fusion rules shown in Table~\ref{Fibotable}
where $0$ corresponds to the vacuum charge $\mathbf{1}$ and $1$ to the Fibonacci anyon $\tau$. Moreover, solving the hexagon and pentagon equations, Eq.~\eqref{hexa} and \eqref{penta}, provides the explicit form for the braid matrix $R$ and the basis change matrix $F$
\begin{equation}
    R_{k=3} = \begin{pmatrix}
    e^{-i\frac{4 \pi}{5}} & 0\\
    0 & -e^{-i\frac{2 \pi}{5}}
    \end{pmatrix},\  \
    F_{k=3} = \begin{pmatrix}
    \varphi^{-\frac{1}{2}} & \varphi^{-\frac{1}{2}}\\
    \varphi^{-\frac{1}{2}} & -\varphi^{-\frac{1}{2}}
\end{pmatrix},
\end{equation}
which in turn can be used to obtain the braid matrices
\begin{equation}
    \sigma_1 = R_{k=3} = \begin{pmatrix}
    e^{i\frac{4 \pi}{5}} & 0\\
    0 & e^{-i\frac{3 \pi}{5}}
    \end{pmatrix}
\end{equation}
and 
\begin{equation}
    \sigma_2 = (F R F^{-1})_{k=3}= \begin{pmatrix}
    \varphi^{-1}e^{-i \frac{4 \pi}{5}} & -\varphi^{-\frac{1}{2}}e^{-i\frac{2 \pi}{5}}\\
    -\varphi^{-\frac{1}{2}}e^{-i\frac{-2 \pi}{10}} & -\varphi^{-1}
\end{pmatrix},
\end{equation}
where $\varphi = (1+\sqrt{5})/2$ is the golden ratio. The emergence of this particular number is due to the expansion of the anyon Hilbert space as anyons are added to the system. Two anyons yield a Hilbert space of dimension 2, three of dimension 3, four of dimension 5 etc., resulting in the Fibonacci sequence and the quantum dimension $d_{k=3} =(1+\sqrt{5})/2$ of the Hilbert space. The quantum dimension may also be computed by means of Eq.~\eqref{qnum} by inserting $k=3$. The Fibonacci generators are $10'th$ roots of unity and the entries of the matrices generated by Fibonacci braids are all members of the polynomial ring $\mathbb{Z}[e^{i\frac{\pi}{5}}] = \{c_1 + c_2 e^{i \frac{\pi}{5}} + c_3 e^{i \frac{2 \pi}{5}} + c_4 e^{i \frac{3 \pi}{5}} | c_i \in \mathbb{Z}\}$. Rings, like groups, are closed structures, which implies that only unitary matrices with entries in this ring can be approximated exactly. 

The ${\rm SU}(2)_2$ Ising anyon model has charges $j = 0, \frac{1}{2}, 1$, where in addition to the vacuum ($j = 0$), we have the Ising anyon $\sigma$ ($j = \frac{1}{2}$ ) and the $\psi$ particle ($j = 1$). Computing the fusion rules of these particles yields the fusion table in Table \ref{isingtable}. Similarly to the Fibonacci model the $F$ and $R$ matrices can be computed to be
\begin{equation}
    R_{k=2} = e^{i\frac{\pi}{8}}\begin{pmatrix}
    -1 & 0\\
    0 & -i
    \end{pmatrix},\  \
    F_{k=2} = \frac{1}{\sqrt{2}}\begin{pmatrix}
    1 & 1\\
    1 & -1
\end{pmatrix},
\end{equation}
which in turn yield the braids matrices
\begin{equation}
    \sigma_1 = R_{k=2} =e^{i\frac{\pi}{8}}\begin{pmatrix}
    -1 & 0\\
    0 & -i
    \end{pmatrix}
\end{equation}
and 
\begin{equation}
    \sigma_2 = (F RF^{-1})_{k=2} = \frac{e^{-i \frac{4 \pi}{8}}}{\sqrt{2}} \begin{pmatrix}
    1 & i\\
    i & 1
\end{pmatrix}.
\end{equation}

An alternative method of determining the $F$ and $R$ matrices is to employ the Temperley--Lieb algebra, which may be regarded as a pictorial string representation of Eq.~\eqref{YBeqns}, and the Jones--Wenzl projectors \cite{wang2010topological,delaney2016local}. This approach allows computation of the matrices for any value of $k$ by means of graphical calculus \cite{wang2010topological}.

It is evident in the structure of the $F$-matrix that the quantum dimension of the Ising anyon model is $d_{k=2}=\sqrt{2}$. An interesting property of the Ising braid group representation for $n=2$ is that it maps bijectively onto the Clifford group, meaning that Ising anyons are an implementation of the Clifford group via braiding. For systems comprised of $n \neq 2$ qubits, however, it only maps onto sub-groups of the Clifford group \cite{ahlbrecht2009implementation}. The Clifford group can be made universal by adding a $\frac{\pi}{4}$-gate, so it must also hold true that the two qubit Ising anyon model becomes universal with this choice of supplementary phase. In Sec.~\ref{sec:results} we will discuss the fact that any non-universal anyon model can be made universal via the addition of an irrational phase gate to the generator set. 
\subsection{Model-$k$ ($k=4,5,6,8$) anyons}

We will briefly outline the structure of the anyon models labelled by $k=4,5,6,8$. Contrary to the Ising and Fibonacci cases, the higher level models provide multiple options of fusion products that can be used as a qubit. For $k=4$ the model contains the charges $j = 0, \frac{1}{2}, 1, \frac{3}{2}, 2$ and the fusion table is provided in Table~\ref{k4table}.

Similarly, Eq.~\eqref{frule} may be used for obtaining the fusion tables for $k=5,6,8$ as well. In the $k=5$ model we have the charges $j=0, \frac{1}{2}, 1 , \frac{3}{4}, 2, \frac{5}{2}$ but due to the duality defined by Eq.~\eqref{autom}, there exist an automorphism from the integer sub-set to the half integer sub-set, which means that we only have to consider one of them. Therefore, in the $k=5$ model it is sufficient to consider the case $j=0,1,2$. The fusion table for $k=5$ anyon model is shown in Table \ref{k5table}.

In the $k=6$ and $k=8$ models such a map between the sub-structures does not exist, which means that all charges are distinct, i.e. $j=0, \frac{1}{2}, 1 , \frac{3}{4}, 2, \frac{5}{2}, 3$ for $k=6$ and $j=0, \frac{1}{2}, 1 , \frac{3}{4}, 2, \frac{5}{2}, 3, \frac{7}{2}, 8$ for $k=8$. Inspecting the fusion tables shows that all of them contain indefinite fusion products, which means that all of the models have some charges transforming under a non-abelian braid group. However, due to Theorems \rom{2.1} and \rom{2.2}, the models labelled by $k=2$ (Ising) and $k=4$ are non-universal, which means that their generator sets must be supplemented with an additional gate in order to carry out general purpose computation. In order to assess the potential performance of various level $k$ anyon models we need a compiler algorithm that maps the unitary operators that realise the desired quantum computation onto the braids of anyons.

\begin{figure}[!b]
    \centering
    \includegraphics[width=0.75\linewidth]{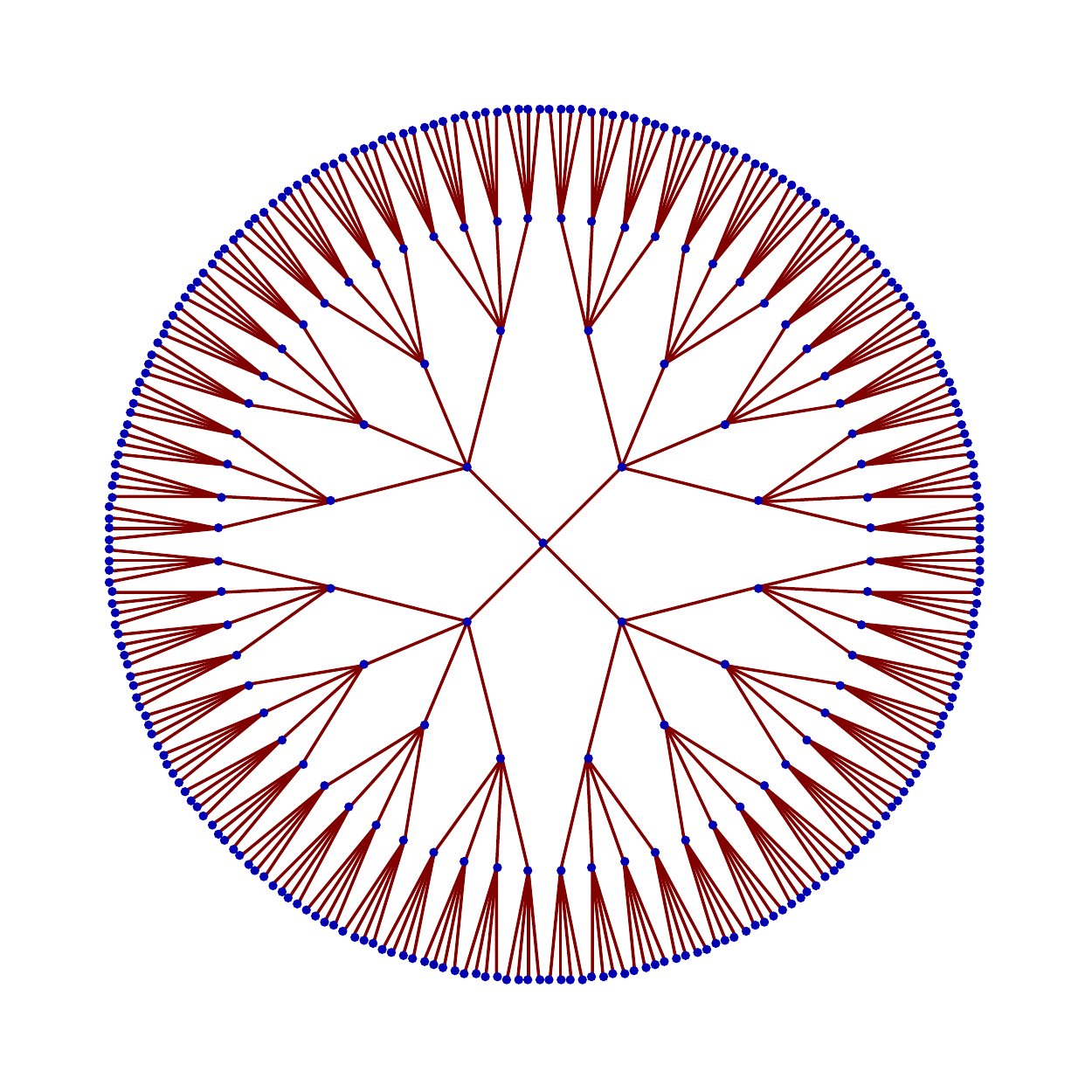}
    \caption{A graphical representation of the search space. The blue nodes correspond to elementary braid operations and the red edges joining the nodes form $4^h$ possible paths at the height $h$ in the search tree. The centre node of the tree corresponds to identity operator.}
    \label{4aryround}
\end{figure}

\section{Quantum Compilation}
\label{sec:quantum compiling}
When a circuit model of computation is carried out in a quantum computer, the state of the system is acted upon by a series of logic gates. These logical operations are induced by physical processes in the system. In order to realize a particular logic gate, we need to know what action it corresponds to in the physical hardware. In TQC, these transformations are realized by braiding the anyons in a very particular way, which raises the question how to deduce what braids should be performed in order to realize the desired computation. A whole research area centered around this problem, known as \emph{quantum compiling}, exists within the field of quantum computation. The key objective of quantum compiling is to develop an algorithm that, given a target unitary gate, will search for sequences of braid operations that are approximating this gate to the desired accuracy. 

\subsection{Exhaustive search and other methods}

\begin{figure}[!b]
    \centering
    \includegraphics[width=\columnwidth]{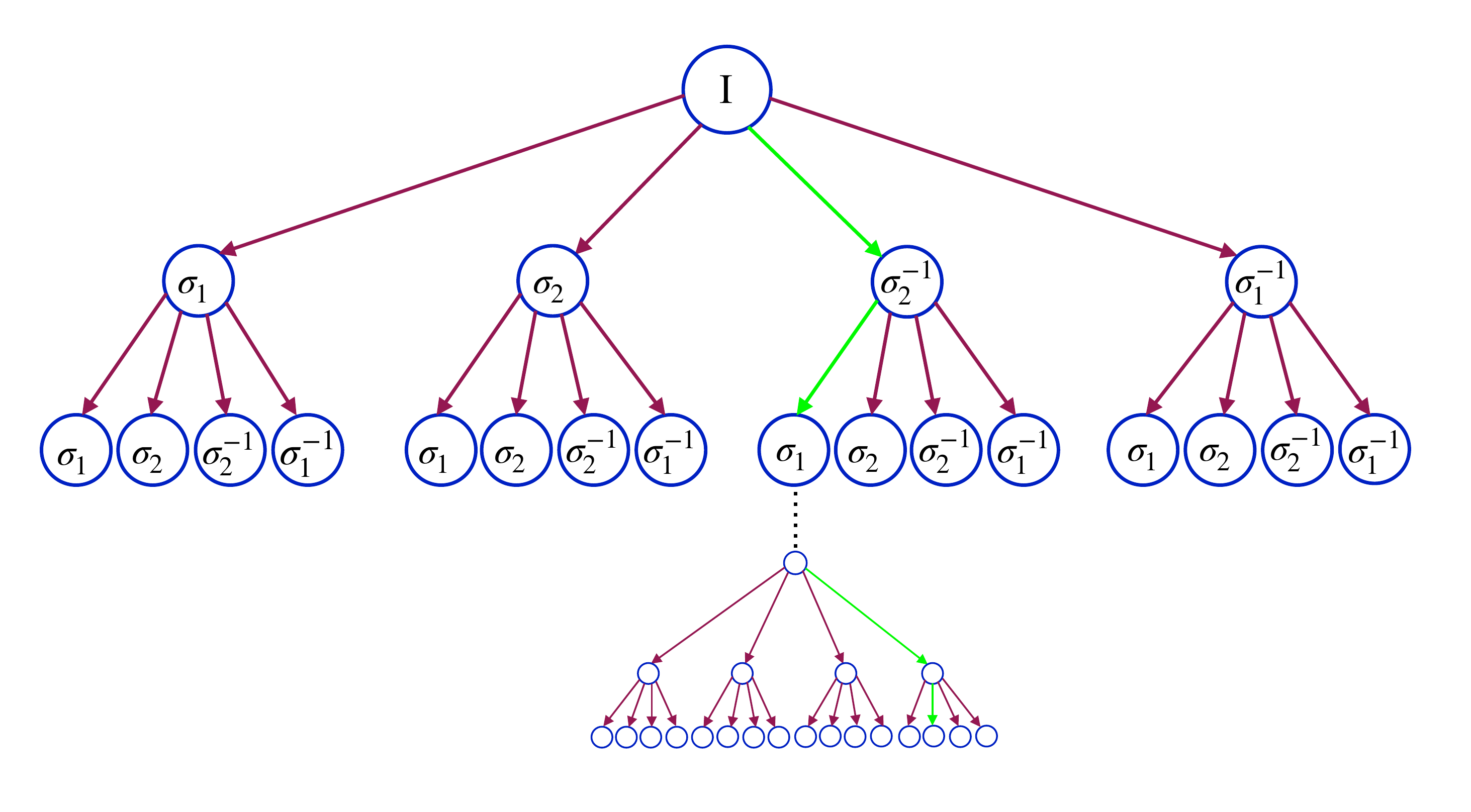}
    \caption{The search space with node labels. The particular  path in green corresponds to the braid word $\mathcal{S}=\mathbb{I}\sigma_2^{-1}\sigma_1 \dotsb \sigma_1^{-1} \sigma_2$.}
    \label{4ary}
\end{figure}

Due to the combinatorial nature of the search problem, we may represent the search space as a tree structure. In particular, if we consider the braid group on three strands, since there are four generators in the image of the corresponding representation this tree is a 4-ary tree shown in Figs~\ref{4aryround} and \ref{4ary}. Considering this graphical structure, the exponential growth of the number of leaf nodes with the length of the braid word (height $h$ of the tree) becomes evident. More precisely, given a tree height $h$, there are a total of $4^h$ leaf nodes. Let us now consider the tree in the context of quantum compilation.

The goal is to find a sequence of elementary braids $\mathcal{S}$, such that the distance $d=||\mathcal{S}-\mathcal{T}||$ to the target unitary matrix $\mathcal{T}$ is less than $\epsilon$, where $\epsilon$ is the desired error tolerance. More specifically, in order to compare the various compiler algorithms as well as anyon models, we need to establish a notion of metric in the space of unitary gates. When a unitary matrix $U$ is approaching a unitary target matrix $U_0$, the product $U_0 U^{\dagger}$ approaches identity such that the error distance $d$ goes to zero. Hence the trace of this product is a number that approaches $n$, where $n$ is the dimension of the matrix. This motivates the definition of the fidelity measure in the space of $2 \times 2$ unitary matrices \cite{kliuchnikov2014asymptotically,field2018introduction}
\begin{equation}
    d(U_0,U) = \sqrt{1-\frac{|Tr(U_0 U^{\dagger})|}{2}}.
\end{equation}
This function has global phase invariance, i.e. the value of the function for a given target matrix $U_0$, is the same for all matrices $U$ that are equal up to a global phase change. It is also strictly positive, symmetric (i.e. $d(V,W)=d(W,V)$) and satisfies the triangle inequality $d(V,W) \leq d(V,U) + d(U,W)$. Each sequence of braids corresponds to a unique path in the search tree, so instead of regarding the problem purely as a constrained optimization problem, we are interested in finding the path to a specific leaf node in the tree, such as the path highlighted in green in Fig.~\ref{4ary}. Note that for an arbitrary $N$-ary tree, given a level $l$, there are a total of $N^l$ leaf nodes. Hence, we may express the total number of nodes in the tree as a geometric sum
\begin{equation}
1+N+N^2+...+N^h = \sum_{l=0}^{h} {N^l} = \frac{N^{h+1}-1}{N-1}.
\end{equation}

Clearly, finding the one leaf node that leads to the smallest error distance $d$ is an inherently exponentially complex problem, which would demand an enormous amount of computational resources. Although performing an exhaustive search is the only known method for finding the unique path to a specific leaf node, good alternative approaches exist. It is possible to reduce the vast size of the search space by considering the algebraic properties of the braid generators and the intra-braid symmetries \cite{field2018introduction}. Finding redundancies in the search space allows pruning whole branches from the tree. There are also techniques for optimizing the brute force search such as the GNAT (Geometric Near-neighbor Access Tree) approach \cite{trung2012optimising}. Methods that are not based on exhaustive search have also been  proposed, e.g. hash function techniques \cite{burrello2010topological} and algebraic techniques \cite{kliuchnikov2014asymptotically, dawson2005solovay}. The method developed in \cite{kliuchnikov2014asymptotically} is a number theoretic approach in which the ring structure of the set generated by the elementary gates is identified, and from that an approximation to a given target gate is constructed. This method is very powerful but not generic in the sense that it would work for any gate set, since it relies on the structure of the generator set. The algorithm discussed in \cite{dawson2005solovay} is known as the Solovay--Kitaev algorithm, which is a scheme that is exploiting the properties of ${\rm SU}(2)$ to recursively achieve better approximations. Contrary to the number theoretic approach due to Kliuchnikov, the Solovay--Kitaev compiler is completely generic although less efficient. We will return to the details of the Solovay--Kitaev algorithm in Sec.~\ref{SK}, after a discussion on Monte Carlo techniques.

\subsection{A Monte Carlo approach}

Let us consider the 4-ary tree shown in Fig.~\ref{4aryround}. Each path in the tree corresponds to a particular braid word. Let $\Sigma$ denote the set of generators and $|\Sigma|$ the order of that set. This implies that for a simple random walk, under the condition that the tree is directed, the probability of moving from a parent node $n_0$ to a specific child node $n_i$, where $i$ is running from $1$ to $|\Sigma|+1$ (number of child nodes plus one parent node), is 
\begin{equation}
P(n_0 \rightarrow n_i) =
\left\{
	\begin{array}{ll}
		\frac{1}{|\Sigma|}\,\  & \mbox{if } n_i = \textrm{child node} \\
		0\,\ & \mbox{if } n_i= \textrm{parent node}.
	\end{array}
\right.
\label{randprob}
\end{equation}

With this particular definition we are only considering directed trees as the probability of going backwards is $0$. We would then like to find the path from the root $\mathbb{I}$ to some node $n$ at level $l$. Thus, if we simulate a simple random walk according to the probability distribution defined in Eq.~\eqref{randprob}, the probability of finding this node is 
\begin{equation}
   P(I \rightarrow n) = \underbrace{\frac{1}{|\Sigma|} \frac{1}{|\Sigma|} \dotsb \frac{1}{|\Sigma|}}_{\text{l \ \ times}} = \frac{1}{|\Sigma|^l}.
\end{equation}

In this model, no move is more favourable than any other, which obviously is a quite simplistic assumption. A better model would be to condition the probability distribution with respect to the present state. Mathematically we can formulate this as a Markov chain from node $n_i$ to node $n_j$ with corresponding states $s_i$ and $s_j$
\begin{equation*}
   P(n_i \rightarrow n_j) =  P(a_k|s_{j-1}) P(s_{j-1})=
\end{equation*}
\begin{equation}
    P(a_k|s_{j-1}) P(a_q|s_{j-2}) P(s_{j-2})=
\end{equation}
\begin{equation*}
     P(a_k|s_{j-1}) P(a_q|s_{j-2}) P(a_p|s_{j-3}) \dotsb P(s_{0}),
\end{equation*}
where we have used the Bayesian property $P(A) = P(A|B)P(B)$. The probabilities $P(a_i|s_j)$ can be interpreted as the probability of taking the action $a_i$, given the present state $s_j$. The remaining problem is how to best model these conditional probabilities.

\subsubsection{Mapping the problem onto a $\mathbb{Z}_N$ spin chain}

\begin{figure}[!b]
    \centering
    \includegraphics[width=0.9\columnwidth]{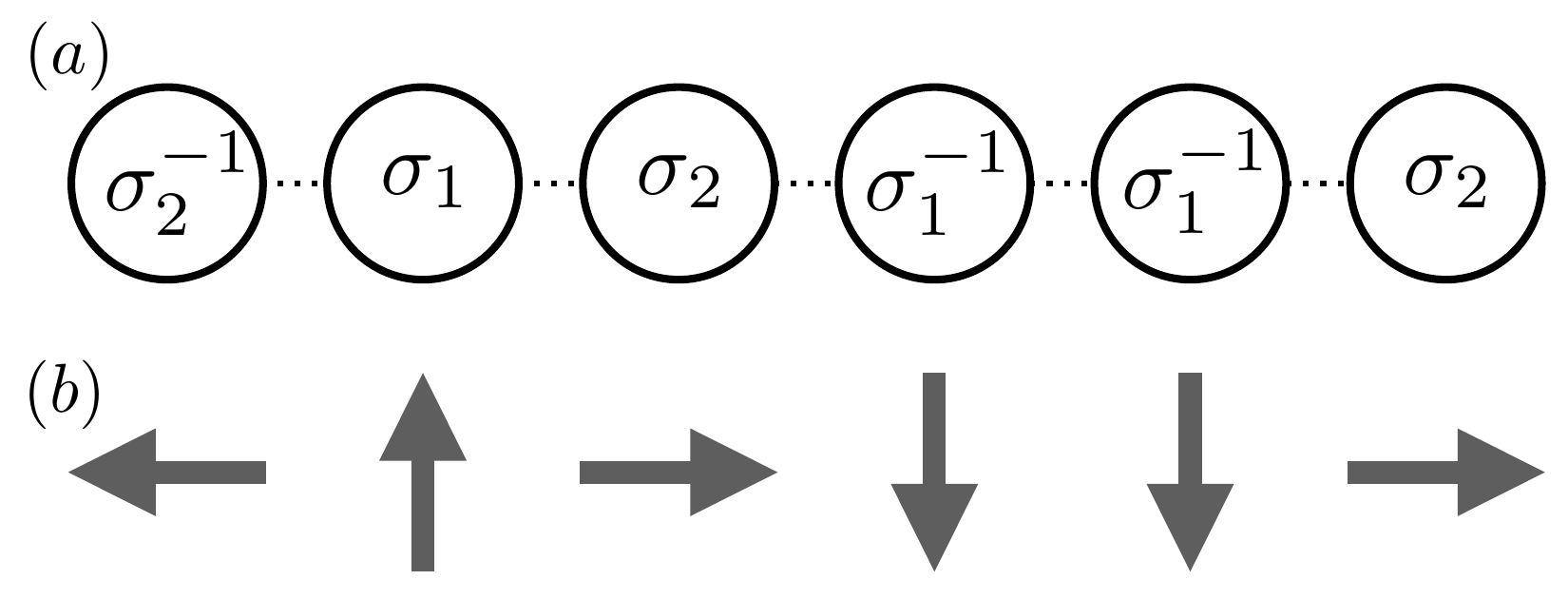}
    \caption{Mapping between a braid word (a) and a spin chain (b).}
    \label{braidspin}
\end{figure}

The core idea of our approach is that the problem of finding the right path in the tree can be mapped onto a one dimensional spin chain for which we are searching for the ground state (see Fig.~\ref{braidspin}). First, let us assign to each generator in the set $\Sigma$ a spin. Here we will consider a set of four generators $\Sigma  = \{ \sigma_1, \sigma_2, \sigma_2^{-1}, \sigma_1^{-1} \}$, which generate the tree in Fig.~\ref{4aryround}. We define the following map $f: \Sigma \rightarrow \mathbb{Z}_4$, 
\begin{equation}\label{pathtospin}
f(\sigma)=
\left\{
	\begin{array}{ll}
		\uparrow \,\  & \mbox{if } \sigma = \sigma_1 \\
		\downarrow \,\ & \mbox{if } \sigma= \sigma_1^{-1}\\
		\rightarrow \,\ & \mbox{if } \sigma= \sigma_2\\
		\leftarrow \,\ & \mbox{if } \sigma= \sigma_2^{-1}.
	\end{array}
\right.
\end{equation}
Thus, for a generic anyon model with $|\Sigma|$ number of generators, we can map this problem onto a $\mathbb{Z}_{|\Sigma|}$ lattice model.

\subsubsection{A thermodynamic picture}

\begin{figure}[!t] 
    \includegraphics[width=\columnwidth]{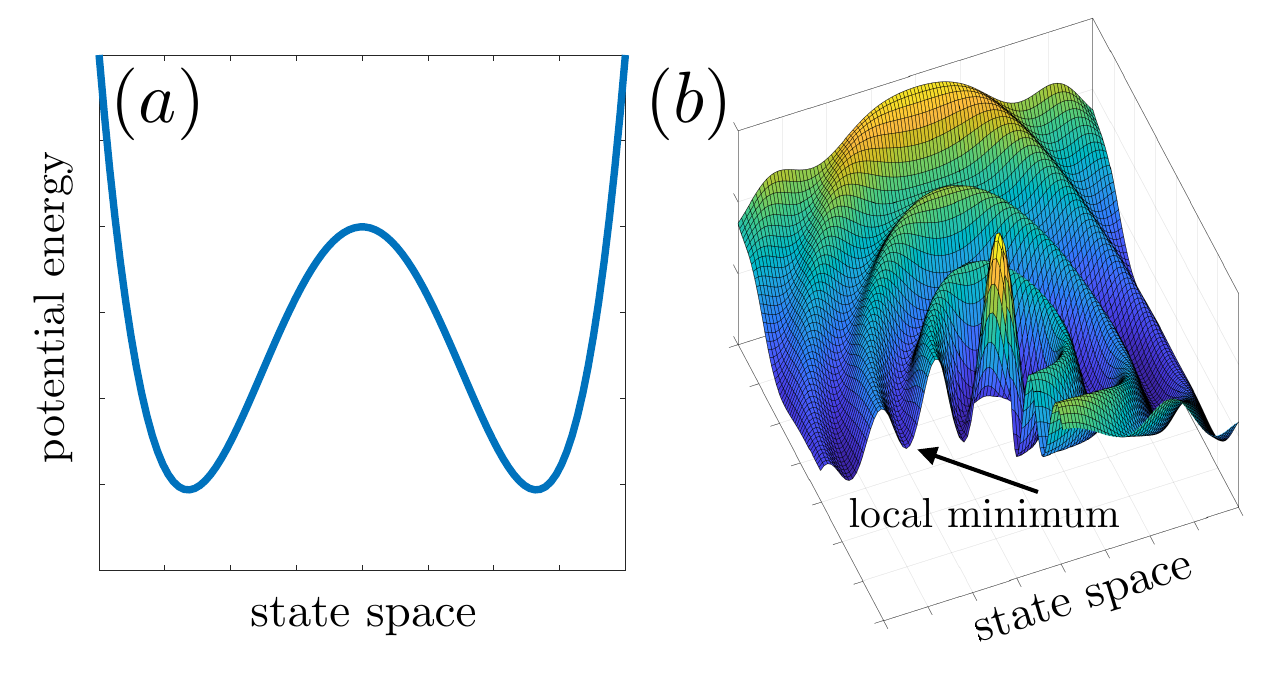}
    \caption{Schematic of the two-fold ground state degeneracy of the Ising anyon model (a) and  a ``wrinkled" Mexican hat potential (b) with several discrete symmetries corresponding to various minima instead of the full U(1) symmetry.}
      \label{isinggroundf}
\end{figure}

Our ultimate goal is to find a path $\mathcal{S}$ corresponding to a braid word that, given a metric, is minimizing the distance $d$ to the target unitary gate $U_0$. We may establish the equivalence $ d \longleftrightarrow E$ in the spin chain representation, where $E$ is the energy of the system. Thus we have implicitly defined a Hamiltonian $H$ of our spin chain via $E=\langle H\rangle$.  Instead of trying to find the smallest error we are interested in finding the state vector corresponding to the known ground state energy of the system. In comparison to the standard Ising lattice model, our implicitly defined spin chain Hamiltonian will most likely exhibit other symmetries. The two-fold degeneracy of the ground state in the Ising spin model is illustrated in Fig.~\ref{isinggroundf}(a) and can be viewed as a 2D version of a Mexican hat function, which is invariant only under the $\mathbb{Z}_2$ subgroup of the full U(1) symmetry. In our scenario the function will look more like a wrinkled Mexican hat function, Fig.~\ref{isinggroundf}(b), which exhibits various discrete symmetries corresponding to a number of local minima. Therefore, in the thermodynamic picture, the error convergence process may be regarded as a series of successive symmetry breaking processes, gradually bringing the system to lower and lower energies. 

Ideally we would like to deduce all of the symmetries as that would allow us to reduce the information entropy, but since these symmetries depend explicitly on the target matrix it is difficult to derive any generally applicable results. This is where the Monte Carlo approach is useful. From the thermodynamic perspective, we would ideally like to find the global minimum of the system, or at least a local minimum that lies close to the global one. Therefore, when the system gets trapped in a local minimum we need to be able to repeatedly climb over barriers and roll down toward lower energy local minima, until the global one is found. This leads to the question of how to assign probabilities to the various minima. In a similar way as for a gas of molecules, we want to maximize the number of statistically equivalent microstates, which happens when the entropy takes its maximum value. Therefore we may introduce a notion of entropy in the system. If we denote by $p_i$ the probability of finding a minimum $i$, the entropy can be expressed as
\begin{equation}
    S = - \alpha \sum_{i=0}^{n} p_i \ln(p_i),
\end{equation}
where $\alpha$ is a constant. The maximum occurs when $\frac{dS}{dp_i} = 0$, and upon imposing stationarity $d(\langle E_i \rangle) = d(\sum_{i=0}^n p_i E_i)= 0$, yields the stationary distribution 
\begin{equation}
 p_i \sim e^{-E_i/T_\alpha},
\end{equation}
where we have introduced a dimensionless temperature parameter $T_\alpha\propto \alpha$ to retain the connection to thermodynamics.

\subsubsection{A Monte Carlo algorithm}

Inspired by the Ising spin model, we assign the event of a spin flip, such as $\uparrow$ flips to $\downarrow$, a probability 
\begin{equation}
 p(\ \uparrow \  {\textrm{to}}  \ \downarrow \ ) = e^{-(E_{\uparrow}-E_{\downarrow})/T_\alpha},
 \label{boltzprob}
\end{equation}
where $E_{\uparrow} > E_{\downarrow}$. In this particular case, we are dealing with four generators and thus four different possible actions, that is, either stay in the initial state, or flip the spin to any of the other three states. The Monte Carlo game proceeds via iterative sequence of attempted and accepted spin flips executed according to the specific rules:
\begin{enumerate}
\item The process  is initiated by performing a simple random walk in the tree according to the probability of Eq.~\eqref{randprob}, to generate an initial spin configuration (braid word) and then moving to the first site (letter). 

  \item A new state of the spin is chosen randomly and if the new state of the system after an attempted flip corresponds to a lower energy, that is $E_{\rm flip}<E_{\rm initial}$, the flip is accepted and the process moves on to the next adjacent site (letter), towards the leaf node.
  
  \item If the energy of the new attempted state is greater, then the flip is accepted with a probability $p$ defined in Eq.~\eqref{boltzprob}. If the flip was not accepted, then another attempt is made at the same site. This procedure is repeated until a flip is either accepted or all possible flips have been attempted once, after which the algorithm proceeds on to the next site.
  
  \item Once the end of the chain (a leaf node) is reached, return to the first lattice site (the root node) and continue the iteration until desired error tolerance is reached.
\end{enumerate}

This basic Monte Carlo method could likely be further optimised by incorporating enhancements such as the worm algorithm \cite{prokof2001worm,boninsegni2006worm} and other stochastic Hamiltonian approaches \cite{bringewatt2019polynomial,jordan2010quantum,bravyi2006complexity}.

\subsubsection{Performance of the Monte Carlo algorithm}

\begin{figure}[!t]
    \centering
    \includegraphics[width=0.9\columnwidth]{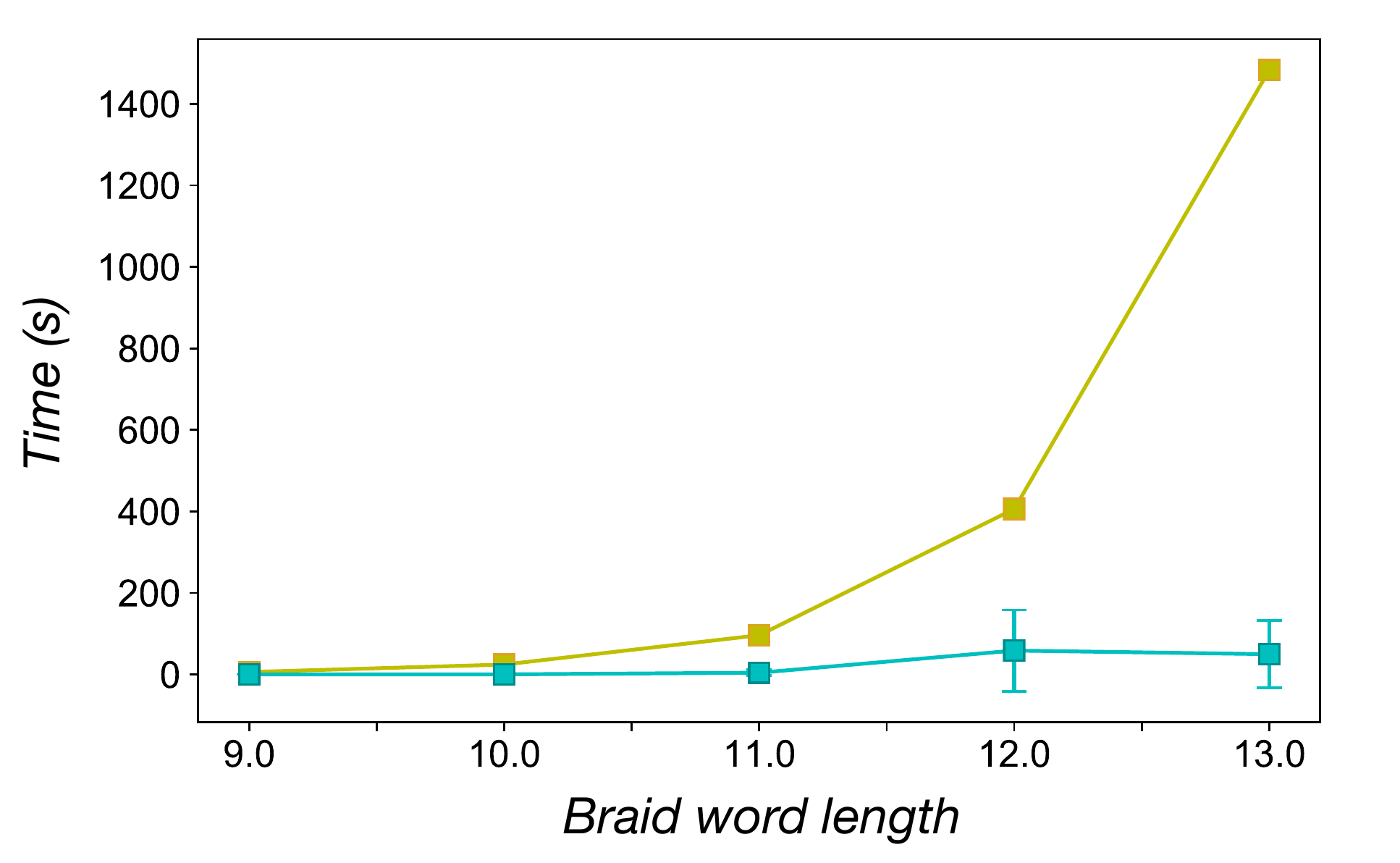}
    \caption{Search time (CPU) as functions of the braid word length for exhaustive search method (yellow) and the Monte Carlo algorithm (turqoise). The errors bars represent the standard deviation of the Monte Carlo sample. The braid words were constructed using the braid generators of the Fibonacci anyon model.
    }
    \label{tvsl}
\end{figure}

The exhaustive search of braid words is always guaranteed to yield the minimum achievable error between the target unitary and its finite-length braid word approximation. It therefore provides a convenient absolute reference for testing the performance of other algorithms. A comparison of the average time taken by a digital computer to find the best possible braid word approximation using an exhaustive search and a Monte Carlo method is presented in Fig.~\ref{tvsl} for a range of braid word lengths within the Fibonacci anyon model.
As expected, the search time grows exponentially with the length of the braid word for the exhaustive search method, which simply reflects the structure of the search space. In  contrast, the Monte Carlo algorithm clearly outperforms the exhaustive search for braid word lengths exceeding 10. 

\begin{figure}[!t]
    \centering
    \includegraphics[width=0.9\columnwidth]{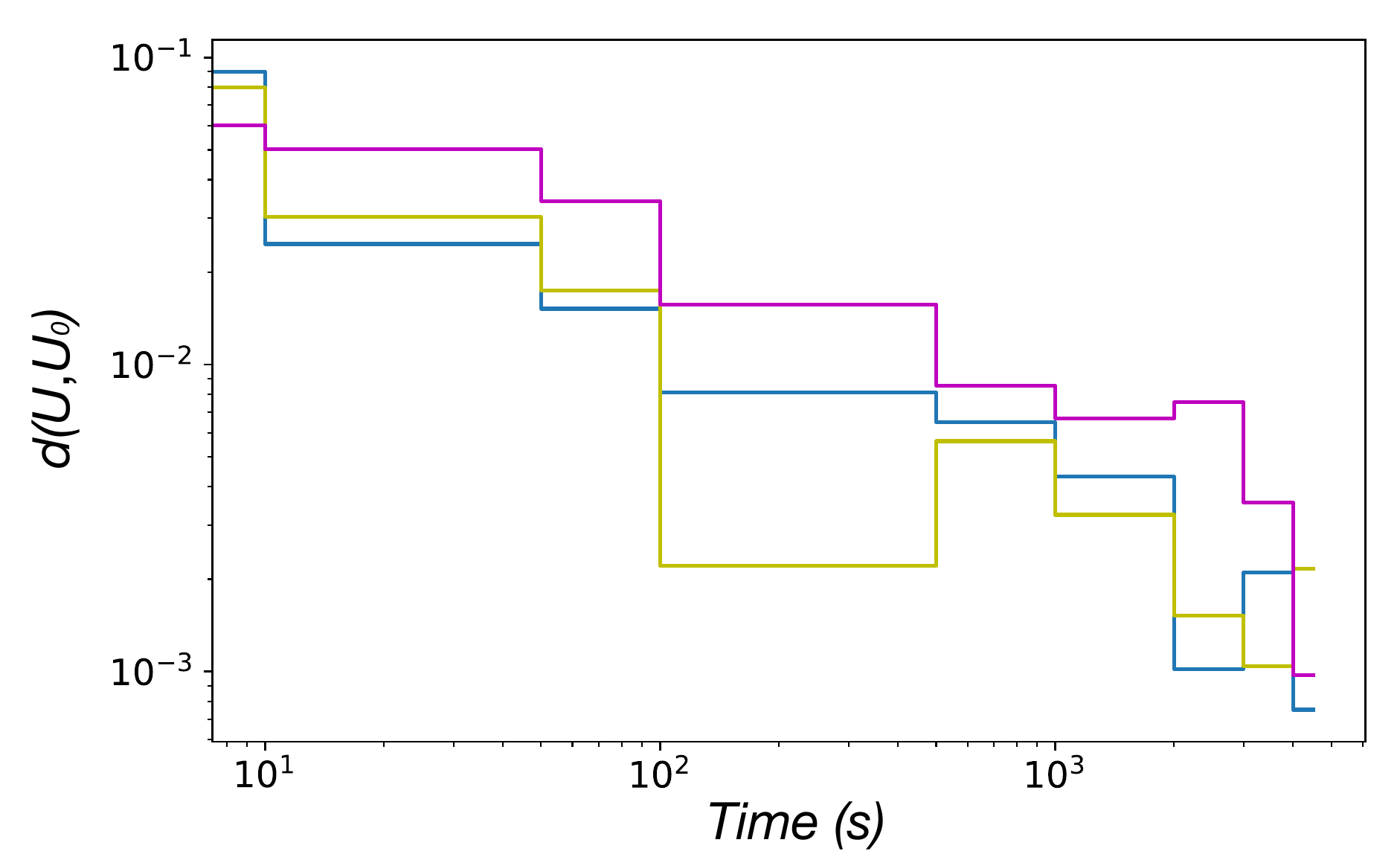}
    \caption{The absolute error $d$ as functions of CPU search time for a Monte Carlo search of a braid word approximation to three randomly chosen unitary matrices. The braidwords of length 50 were constructed using the braid generators of the Fibonacci anyon model.}
    \label{evst}
\end{figure}

To understand the convergence properties of the Monte Carlo algorithm, Fig.~\ref{evst} shows the achieved absolute error as a function of CPU time for three distinct randomly generated unitary target gates and a braid word length of 50.
Figure~\ref{evst} shows the overall convergence behaviour of the error. Occasionally, the algorithm gets stuck in a local minimum corresponding to the plateaus in the figure, but eventually `tunneling' over the barrier is achieved and the converge proceeds toward another local minimum. 
The Monte Carlo algorithm is particularly useful when compiling braids for computationally non-universal models such as the Ising anyon model that require at least one additional elementary gate to be added to the generator set in order to make the model computationally universal. This means that the search space grows as $5^l$ instead of $4^l$, which makes it even more challenging to perform an exhaustive search. More importantly, the additional gate is not topologically protected and is thus susceptible to ordinary forms of decoherence. It is therefore of great interest to reduce the total number of such non-topological gates required to approximate a predetermined target unitary. As shown in Sec.~\ref{noise}, it is straightforward to implement this within the Monte Carlo algorithm by introducing a phase gate acceptance probability. If we define the probability such that the probability is decreasing with the number of phase gates already accepted, it is possible to reduce the phase gate count significantly and thus suppress the susceptibility of the computation to conventional forms of decoherence.

\subsection{The Solovay--Kitaev algorithm}
\label{SK}
The Solovay--Kitaev algorithm can be viewed as an algoritmic implementation of the shrinking lemma that leads to the Solovay--Kitaev theorem. The algorithm plays a significant role in quantum computation as it guarantees a converging sequence of gates approximating any target gate can be found with any desired accuracy. Here we will follow Dawson and Nielsen's implementation in Ref.~\cite{dawson2005solovay}.

\subsubsection{The Shrinking lemma and the Solovay--Kitaev theorem}

Let $G$ be a set of generators forming a space endowed with a metric $d(X,Y)$. For any accuracy $\epsilon>0$, there is a constant $c$ and a sequence of generators $S$ of length $O(\log_c (\frac{1}{\epsilon}))$, such that the distance to a given target matrix U satisfies $d(S,U) < \epsilon$, if the set is dense in ${\rm SU}(2)$. This statement constitute the basis upon which the Solovay--Kitaev algorithm is founded. In simple terms the shrinking lemma states that by recursively expanding the length of the braid word, it is possible to get closer and closer to the target gate.

In order to be able to realise any arbitrary gate, an anyon model must be \emph{universal}. In terms of the Bloch sphere representation, this implies that it is always possible to find a combination of generators such that their joint action on the state vector maps the initial point on the sphere arbitrarily close to any other point. Algebraically, this entails that ${\rm SU}(2)$ must be entirely contained in the set, or that each point in ${\rm SU}(2)$ has a limit point in the set that is arbitrarily close to that point, given that the space is endowed with a metric. It is thus stated that the set is topologically dense in ${\rm SU}(2)$ or that the set provides a dense cover for ${\rm SU}(2)$. This mathematical statement formalises the Solovay--Kitaev theorem. For a more detailed analysis, see e.g. \cite{dawson2005solovay, nielsen2002quantum, kitaev1997quantum}. The Solovay--Kitaev theorem has important implications to quantum computation since it ensures that universal quantum computation is indeed possible, at least in theory. It was later shown that the bound on the braid word length to error ratio could be improved, leading to a plethora of new adaptations. However, the improvements discussed in the literature are non-generic, that is, they only hold for specific generator sets \cite{kliuchnikov2013synthesis, paetznick2013repeat, selinger2012efficient, bocharov2015efficient, ross2014optimal, kliuchnikov2015practical, paetznick2013repeat}. The analysis in the following sub-section is mainly based on Ref.~\cite{nielsen2002quantum}.

\subsubsection{Implementation of the Solovay--Kitaev algorithm}

 The key step in the Solovay--Kitaev algorithm, and also in the proof of the shrinking lemma, is to perform a group commutator decomposition (GCD) of the quantity $\Delta = U U_0$, with $U$ being the approximation and $U_0$ the target gate. Then, by finding approximations to the factors in the decomposition, their product yields a better approximation of $\Delta$, than what could be obtained by searching for approximations of $\Delta$ directly. When the decomposition is performed, the function calls upon itself recursively with the factors in the decomposition as inputs, and these input matrices are further decomposed to achieve even higher accuracies.
 
Since $U = V W V^{\dagger} W^{\dagger} U_0^{\dagger}$, and each one of the factors in the decomposition is of length $l_0$, the total length of a braid word that is a level-$n$ approximation is given by $l_n = l_0 5^n$. The implementation of Algorithm~\ref{SKA} is due to Dawson and Nielsen \cite{dawson2005solovay}.

\begin{algorithm}[!t]
\SetAlgoLined
  \eIf{$n=0$}{
   braid word = exhaustive search$(U)$\;
   \textbf{return} braid word\
   }{
   $\widetilde{U}$ = Solovay--Kitaev$(U,n-1)$\;
   
   $\Delta = U \widetilde{U} = \widetilde{V} \widetilde{W} \widetilde{V}^{\dagger} \widetilde{W}^{\dagger};$
   
   $(V, W)$ = GCD$(\Delta)$;
   
   $\widetilde{V}$ = Solovay--Kitaev$(V,n-1)$\;
   
   $\widetilde{W}$ = Solovay--Kitaev$(W,n-1)$\;
   
   $U_n = \widetilde{V} \widetilde{W} \widetilde{V}^{\dagger} \widetilde{W}^{\dagger} \widetilde{U}^{\dagger}$;
   
   \textbf{return} $U_n$
  
 }
 \caption{Solovay--Kitaev$(U,n)$}
 \label{SKA}
\end{algorithm}
 
\subsection{Monte Carlo enhanced Solovay--Kitaev algorithm}

The development of the Solovay--Kitaev algorithm was a major step forward in the field of quantum compiling. Nevertheless, it has a few downsides. Probably the most severe is the exponential growth of the braid word length as a function of the level $n$ of the approximation, which due to the group commutator decomposition grows as
\begin{equation}
    l_k = l_0 5^n,
\end{equation}
where $l_0$ is the braid word length of the zeroth order approximation. In order to reach high accuracies, one has to go to great depths (large $n$) in the algorithm and this results in very long braid words. For instance, to reach an accuracy of $\epsilon \sim 10^{-4}$ with a base length $l_0 = 10$, depth $n=5$ may be needed, which means that the length of the resulting braid word is $l_5 = 10 \cdot 5^5 = 31250$, which is unnecessarily long for an error of that magnitude (even though according to the threshold theorem errors below $1\%$ are acceptable in order to perform quantum computation fault-tolerantly \cite{raussendorf2007fault, fowler2009high, fowler2012surface, campbell2017roads, gottesman2010introduction,webster2019fault,webster2018braiding,brown2013topological}). Another downside is the  required computation time. Considering the pseudo code of the previous sub-section, we conclude that the algorithm makes three recursive calls per level. This implies that the simulation time increases by a factor three per level according to
\begin{equation}
    t_n = t_0 3^n,
\end{equation}
where $t_0$ is the time it takes to perform the exhaustive search.  

In summary, while the algorithm promises unconstrained accuracy, the accuracy realised for a given braid word length is highly sub-optimal. With an ideal algorithm, it should be possible to find braid words of this accuracy that are only a fraction of the length. Moreover, the exponential growth in simulation time makes the process very slow. The simplest way to improve the efficiency of the algorithm, is to find a method that enables surpassing the limits of the exhaustive search method. That is, finding better approximations for shorter braid words, in a shorter time. The graphs presented in Figs~ \ref{tvsl} and \ref{evst}, show that the Monte Carlo algorithm is significantly faster than the exhaustive search method. While the exhaustive search method is exponential in time, the Monte Carlo one is linear (at least for moderately long braid words), and since it can easily be applied beyond the small search spaces that the brute force method is limited by, it is also possible to find much better zeroth depth, $n=0$, approximations. In addition to the search time, also the search space (n-ary tree) has to be constructed before performing an exhaustive search. This is an inherently exponential problem in itself as the search space grows as $m^h$, where $m$ is the number of child nodes per node and $h$ is the height of the tree. Thus, we may conclude that the advantage of the Monte Carlo method is two-fold: (i) it enables better zeroth level approximations, and (ii) not only is it searching more efficiently, but also no preparatory work is required prior to the initiation of the search process. Additionally, it is possible to suppress the phase gate count in the non-universal models by introducing a phase gate acceptance probability to the Monte Carlo algorithm, which makes it particularly powerful for hybrid anyon models. More general constraints on the desired braid words are also straightforward to implement within the Monte Carlo method.

By implementing the Monte Carlo method instead of an exhaustive search at the zeroth depth in the Solovay--Kitaev algorithm, it should be possible to enhance the performance significantly. The implementation of this enhancement is very simple. The exhaustive search method at depth $n=0$ in Algorithm \ref{SKA} is simply substituted with the Monte Carlo search.

\subsection{Comparison of compiler algorithms}

\begin{figure}[!t]
    \centering
    \includegraphics[width=90mm,scale=1]{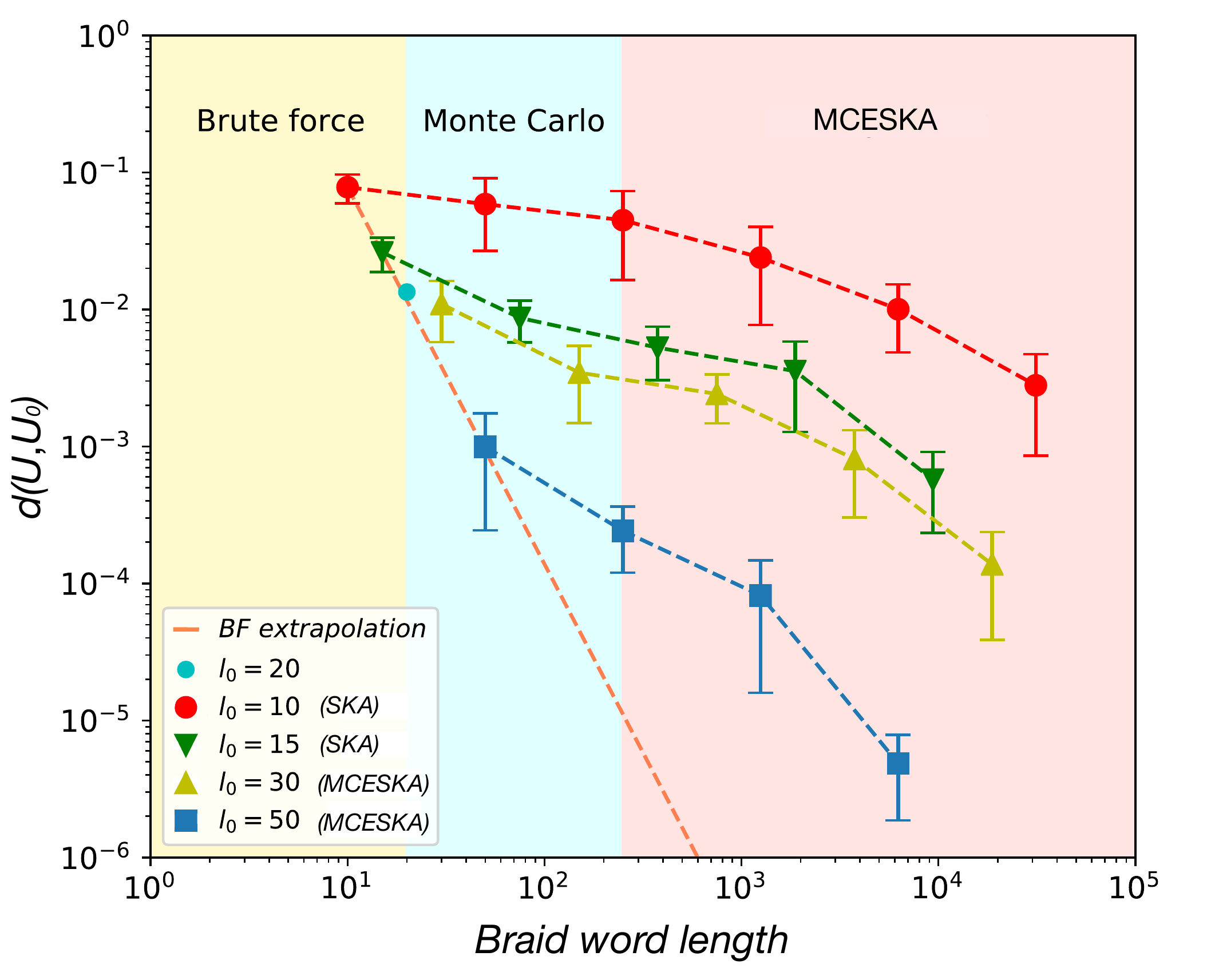}
    \caption{Comparison of the performance of the conventional Soloway--Kitaev quantum compiler algorithm (SKA) with our Monte Carlo enhanced version (MCESKA). The dashed orange line has a slope of $-0.0017275$ and is an extrapolation based on the data points obtained with the brute force method and should be viewed as an estimation of the lowest theoretically attainable error for given braid word length. The results were obtained using the Fibonacci anyon model.}
    \label{compilercomp}
\end{figure}

Figure~\ref{compilercomp} summarises our quantum compiler benchmark results obtained for the conventional Solovay--Kitaev algorithm (SKA) and our Monte Carlo enhanced Solovay--Kitaev algorithm (MCESKA) using the braid generators of the Fibonacci anyon model. We have used $l_0 = 10$ and $l_0 = 15$ as base lengths for SKA and $l_0 = 30$ and $l_0 = 50$ for MCESKA. The dashed orange line is an extrapolatation based on the brute force method data points (red, green and cyan). This is a useful benchmark for how well our algorithm is performing compared to an hypothetical ideal algorithm that would find the best approximation for any braid word length. In particular, the pure Monte Carlo method for $l_0 = 50$ achieves practically equivalent precision as an estimated brute force method would. The shaded regions correspond to the braid word length intervals for which the specified algorithm is recommended to be used for optimal results.

The results in Fig.~\ref{compilercomp}, confirm our conjecture. Invoking the Monte Carlo method at the zeroth depth brings substantial benefits. Note that these results are obtained for particular parameter values in the algorithm, which could be further optimised. For instance, if one succeeds to improve the calibration of the temperature parameter in Eq.~\eqref{boltzprob}, a higher convergence rate could be achieved. Also, if we allow for a longer simulation time, the algorithm will be able to settle into lower local minimum. With this particular set of parameter values we see that the blue curve ($l_0 = 50$) reaches an error two orders of magnitude lower than the green one ($l_0 = 15$), after only three MCESKA recursions. Nevertheless, the braid word length corresponding to this error is still roughly one order of magnitude greater than the estimated ideal length obtained from the brute force extrapolation (the dashed orange line).

\section{Numerical experiments}
\label{sec:results}
\subsection{Fibonacci versus Ising anyon models}
Here we present MCESKA compilation results for the Fibonacci and Ising anyon models. As mentioned previously, the Fibonacci model is capable of universal quantum computation by only braiding the anyons, whereas the Ising anyon model must be supplemented with an additional unitary operator such as a suitably chosen phase gate. It is thus natural to ask how to choose such an auxliary gate?

Recall that universality implies that we must be able to arrive arbitrarily close to any point on the Bloch sphere just by combining the generators. In $\mathbb{R}^2$, by picking any two vectors $\Bar{v},\Bar{w}$, such that $\Bar{v} \times \Bar{w} \neq 0$, and parameters $a,b \in \mathbb{R}$, we may construct a linear combination $(a\Bar{v},b\Bar{w})$ that reaches any point in the plane. Similarly, on the sphere, by choosing a phase gate $\phi$ such that it is possible to construct at least two continuous and non-parallel rotations, we should be able to cover the whole sphere. One way to achieve this is to select a phase that is generating a dense set in $S^1$, since then we can combine this phase with the other generators to form multiple ``non-parallel" dense circles. Mathematically, we thus need to select a phase of infinite order, i.e. $\phi^n \neq I$ for all $n \in \mathbb{Z} \setminus \{0\}$, and find at least two sequences $s_1,s_2$ containing this phase, such that $[s_1,s_2] \neq 0$. By considering the unit circle in the complex plane, the two phases $e^{i n x 2 \pi}$ and $e^{i m y 2 \pi}$, where $m,n \in \mathbb{Z}$, can never be equal if we let $x,y \in \mathbb{R} \setminus \mathbb{Q}$ (the irrationals), since no such number can be expressed as a fraction, which further implies that $e^{i n x 2 \pi} \neq e^{i m y 2 \pi}$ $\forall n \neq m$. Hence, we arrive at the following propositions:
\newline\newline
\textbf{Proposition \rom{5.1}} \\
\textit{Let $\phi = e^{i x \pi}$ be a phase on the complex unit circle. Then the set generated by $\phi ^n$, where $n \in \mathbb{Z}$, will form a topologically dense cover in $S^1$ (in the complex plane), if $x \in \mathbb{R} \setminus \mathbb{Q}$,}\\
\newline
and
\newline\newline
\textbf{Proposition \rom{5.2}} \\
\textit{Let $\Sigma$ be a set of generators that is not forming a topologically dense cover in ${\rm SU}(2)$. Then by adding an irrational phase gate $\phi$, it will become dense if at least two sequences $s_1$ and $s_2$, containing $\phi$, can be found such that $[s_1, s_2] \neq 0$.}\\
\\
Having established that any irrational phase can be used to supplement a non-universal anyon model, we only need to find two sequences containing this phase that do not commute, in order to form a basis on the Bloch sphere. This is easily achieved since braid generators do not commute in general. For instance, $\sigma_1 \phi$ will generate a dense circle around one axis and $\sigma_2 \phi$ around another one since $[\sigma_1 \phi, \sigma_2 \phi] \neq 0$. Thus we may conclude that universality can be achieved in the Ising anyon model by adding an irrational phase gate to the set of generators. An in depth analysis of the conditions for universality can be found in \citep{sawicki2017criteria} and physical implementations of phase gates are discussed in \citep{bonderson2010implementing}. 

In our numerical experiments, we generate a large number of random unitary matrices as target gates that are then compiled using the different anyon models and for a range of braid word lengths. The braid generators for generic level $k$ anyon model considered are constructed using Eqns~(\ref{bg1})-(\ref{sixj}).

\begin{figure}[!t]
    \centering
    \includegraphics[width=\columnwidth]{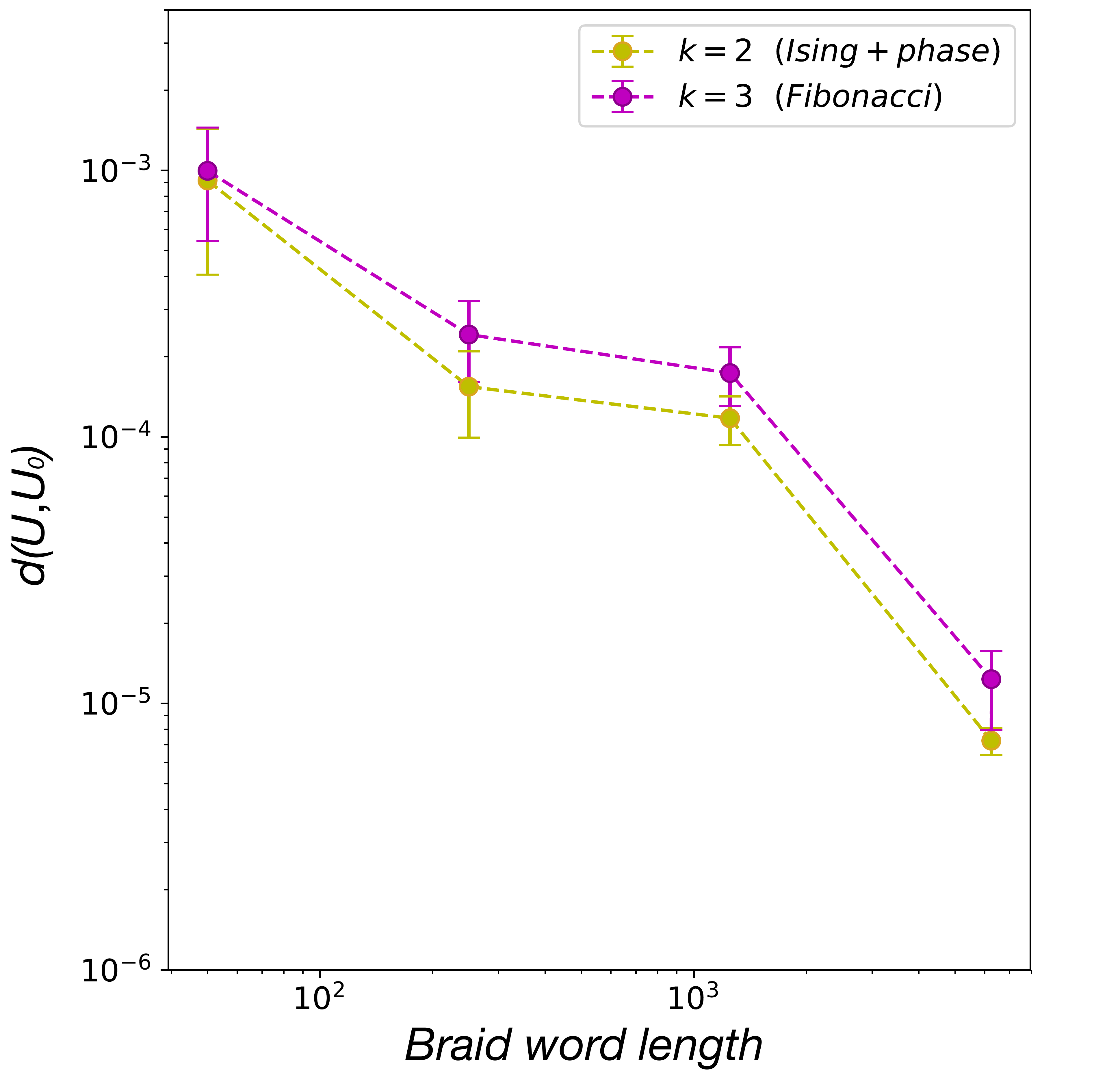}
    \caption{Compilation errors as functions of the braid word length evaluated for the Fibonacci and hybrid Ising anyon models. In total about $20\%$ of $e^{i\sqrt{2}\pi}$ phase gates were deployed in the Ising braid words. Each data point comprises $10^3$ realisations with a corresponding statistical error estimate shown.}
    \label{fibvsising}
\end{figure}

Figure \ref{fibvsising} shows a comparison between the universal Fibonacci anyon model and the hybrid Ising anyon model for which a phase gate 
\[R_\theta =\left(\begin{matrix} 1&0\\0&e^{i\theta} \end{matrix}\right)\]
with a phase $\theta=\sqrt{2} \pi$ was used to achieve universality. The data points in Fig.~\ref{fibvsising} are an average over the sample with one thousand data points and the error bars represent the corresponding standard deviation. 
This numerical experiment confirms our conjecture that the Ising anyon model becomes computationally universal when enhanced with an irrational phase. The average performance of this hybrid Ising anyon model seems generally a slightly better than the Fibonacci model but since the error bars are overlapping, from a statistical point of view, their performance should be considered to be equal. However, since the additional phase gate cannot be implemented in a topologically protected manner, the hybrid Ising anyon model is not immune to decoherence, which means that from the practical point of view the Fibonacci model will always turn out to outperform the hybrid Ising anyon model in the presence of environmental noise, as is discussed in detail in Sec.~V D.

\subsection{The level $k=4,5,6,8$ anyon models\label{subsec:kcomp}}
To provide a broader perspective, we extended our analysis of the $k=2$ Ising and $k=3$ Fibonacci anyon models to the level $k=4,5,6,8$ anyon models. In light of Theorems \rom{2.1} and \rom{2.2}, we conclude that, similar to the $k=2$ case, the $k=4$ model is non-universal and need to be supplemented with a generator having an irrational phase. We used the fusion product $\frac{k-1}{2} \otimes \frac{k-1}{2} = 0 \oplus 1$ as qubit for these numerical experiments.

\begin{figure*}[htp!]
    \centering
    \includegraphics[width= 1.8\columnwidth]{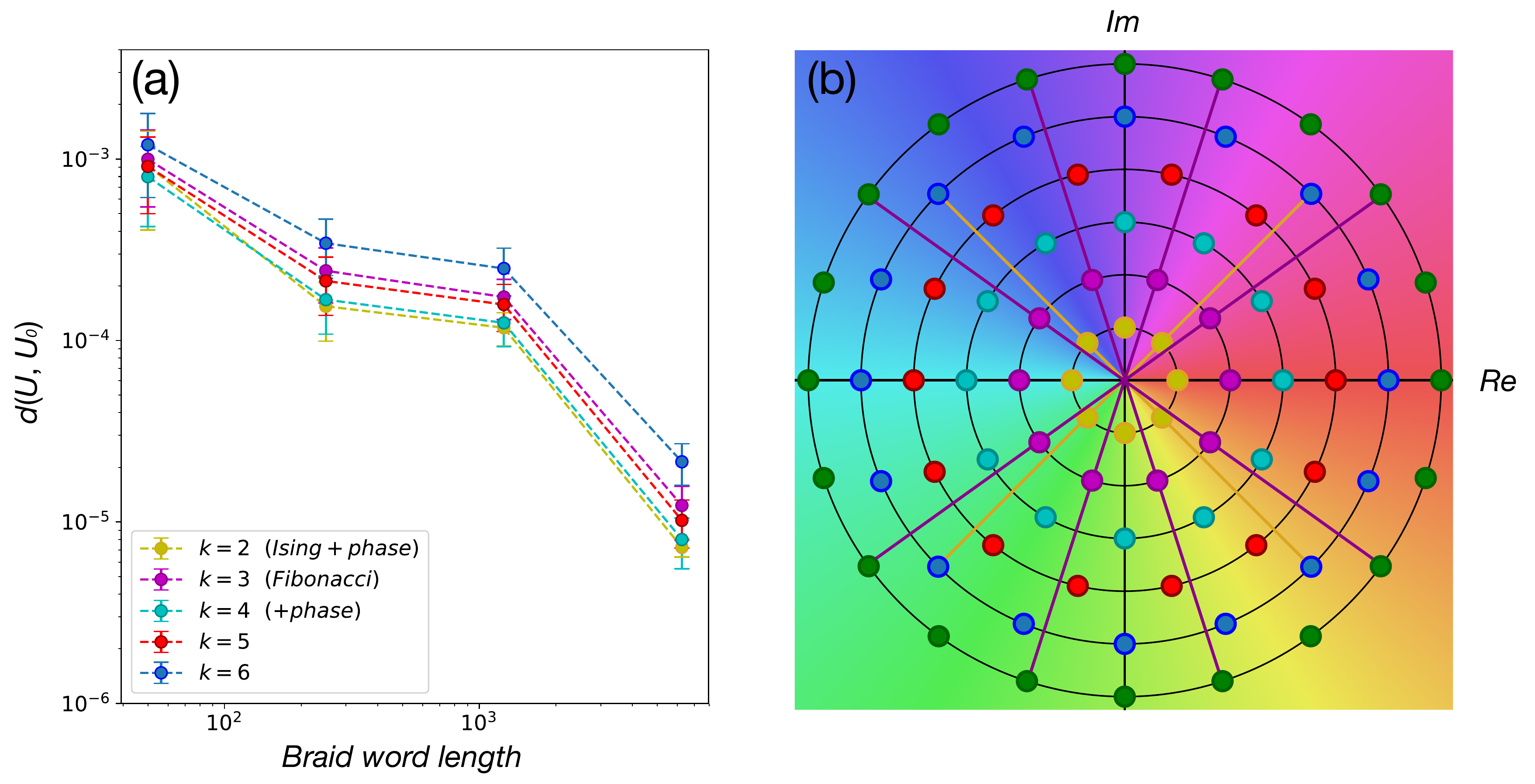}
    \caption{Comparison of level $k=2,3,4,5,6,8$ anyon models. For all models the compilation error as functions of the braid word length (a) is consistent with an overall powerlaw convergence to the target gate. The cyclicity of each model, illustrated on the Argand plane (b), reveals that the $k=2$ model is contained within the $k=6$ model and that the $k=3$ model is contained within the $k=8$ model. The $k=2$ and $k=4$ cases were supplemented with $e^{i\sqrt{2}\pi}$ phase gates. Each data point comprises $10^3$ realisations with a corresponding statistical error estimate shown.
    }
    \label{allk}
\end{figure*}

The results of these calculations are summarised in Fig.~\ref{allk}, which shows that the data corresponding to $k=2,3,4,5,8$ are all within one standard deviation from one another. The only one that lies consistently above the others is the $k=6$ model. This indicates that the group generated by the $k=6$ braid generators may cover the ${\rm SU}(2)$ at a slower rate than the other models, yielding a slower convergence rate. In order to understand this observation, we construct a heuristic argument based on the structure of the corresponding braid group. We will project the rotations onto the complex plane to analyse the cyclicity of the group generators. Letting $\Sigma_{\sigma_i , k}$ denote the sub-group generated by $\sigma_i$ for a given $k$, we may define an isomorphism
\begin{equation}
    f: \Sigma_{\sigma_i , k} \longrightarrow \Sigma_{\sigma_i , k}',
    \label{fmap}
\end{equation}
such that $f(\sigma_i^q \sigma_j^p) = f(\sigma_i^q)f(\sigma_j^p)$ for any $q,p \in \mathbb{Z}$, where the image forms a group that is identical in terms of rotations but the axis is aligned with the axis of the group we are comparing with. In fact, $\Sigma_{\sigma_i , k}$ and $\Sigma_{\sigma_i , k}'$ belong to the same equivalence class. Since the generators $\sigma_1$ and $\sigma_2$ in a given ${\rm SU}(2)_k$ model always are $2(k+2)$-cycles, we can deduce the size of the intersection between two sub-groups $\Sigma_{\sigma_i,k}'$ and $\Sigma_{\sigma_i,k'}$ (where $\Sigma_{\sigma_i,k}'$ is the image of $f$ such that the axes are aligned), generated by $\sigma_i$. Let us start with the case in which $k,k' \in 2\mathbb{Z}$ (even) and let $k'>k$, then the following holds when projected onto the complex plane
\begin{equation}
  \Sigma_{\sigma_i,k}' \bigcap \Sigma_{\sigma_i,k'} =
    \begin{cases}
      \Sigma'_{\sigma_i,k} & \text{, if $k+2$ \textbar $k'+2$}\\
      \{0, e^{i\frac{\pi}{2}}, e^{i \pi}, e^{i \frac{3 \pi}{3}}\} & \text{, otherwise}.
    \end{cases}       
\end{equation}
From this we deduce that $\Sigma'_{k}$ is completely contained in $\Sigma_{k'}$ if $k'+2$ is divisible by $k+2$, whereas if this condition is not satisfied, the intersection will form a smaller group corresponding to $\frac{\pi}{2}$-rotations. Letting $k \in 2 \mathbb{Z} - 1$ (odd) and $k' \in 2 \mathbb{Z}$ (even), the intersection is given by
\begin{equation}
  \Sigma_{\sigma_i,k}' \bigcap \Sigma_{\sigma_i,k'} =
    \begin{cases}
      \Sigma'_{\sigma_i,k} & \text{, if $k+2$ \textbar $k'+2$}\\
      \{0, e^{i \pi}\} & \text{, otherwise}.
    \end{cases}       
\end{equation}
That is, the odd $k$ sub-group is completely contained in the even one if $k'+2$ is divisible by $k+2$, and if not, the intersection is constituted by the group corresponding to $\pi$-rotations. However, the reverse is never true, where the even $k$ group would be completely contained in the odd $k$ one since all even $k$ groups contain a $\frac{\pi}{2}$-and $\frac{3 \pi}{2}$-rotation, while odd $k$ groups never do. 

If we now apply this analysis to our particular case by comparing the $k=6$ group with the others. As depicted in Fig.~\ref{allk}(b), this group is a $16$-cycle at generator level and for all $k=3,4,5$, we have that $k'+2 \nmid k+2$, i.e. an $16-$cycle is not divisible by the $10,12,14$ ones ($k=3,4,5$). However, the generators in the $k=2$ are $8-$cycles and $8\mid 16$, so all rotations in the $k=2$ are being faithfully represented by elements in the $k=6$ braid group. Or equivalently, the structure of the $k=2$ braid group is entirely encoded in the $k=6$ one, which is not true for $k=3,4,5$. Hence, there is no a priori reason to expect the $k=6$ model to perform better than the $k=3,5$ models, since the $k=2$ model is non-universal. In fact, the $k=6$ model is the first even $k$ model that is universal. Moreover, if we apply the same analysis to $k=8$ we find that $k+2$ \textbar $k'+2$ only if we let $k'=3$. This means that the Fibonacci braid group is isomorphic (according to $f$ defined in Eq.~\eqref{fmap}) to a sub-group of the braid group labeled by $k=8$. So in this case it should hold that the $k=8$ model is at least as good as the Fibonacci model with regards to braiding, which also conforms with the results shown in Fig.~\ref{allk}.

\subsection{Decoherence in non-universal anyon models}
\label{noise}
Adding a phase gate to the otherwise topologically protected braid generators comes with a price. Unlike the topological gates that are immune to conventional sources of noise, the added phase gate generator cannot be implemented in hardware without accompanying decoherence. The more phase gates are used, the more adverse the effect of the noise brought along is. It is therefore preferable to keep the phase gate count as low as possible, as the accumulative effect of the noise during braiding may have terminal consequences for the computation.

For the Ising anyon model we might expect the unconstrained fraction of phase gates contained in a braid word to be about 20\% on average for long sequences, since there are in total five generators, four topological and one conventional, in the set. To see if it is possible to suppress the dependence on this ``necessary but undesirable" generator,  we introduced an acceptance probability in the Monte Carlo algorithm $p \sim e^{-c \frac{n}{l}}$, where $c \in \mathbb{R}$, $n$ is the number of phase gates in the sequence and $l$ is the length of the braid word. Thus, as $n$ is growing, the acceptance probability is decreasing.
\begin{figure}[!b]
    \centering
    \includegraphics[width=\columnwidth]{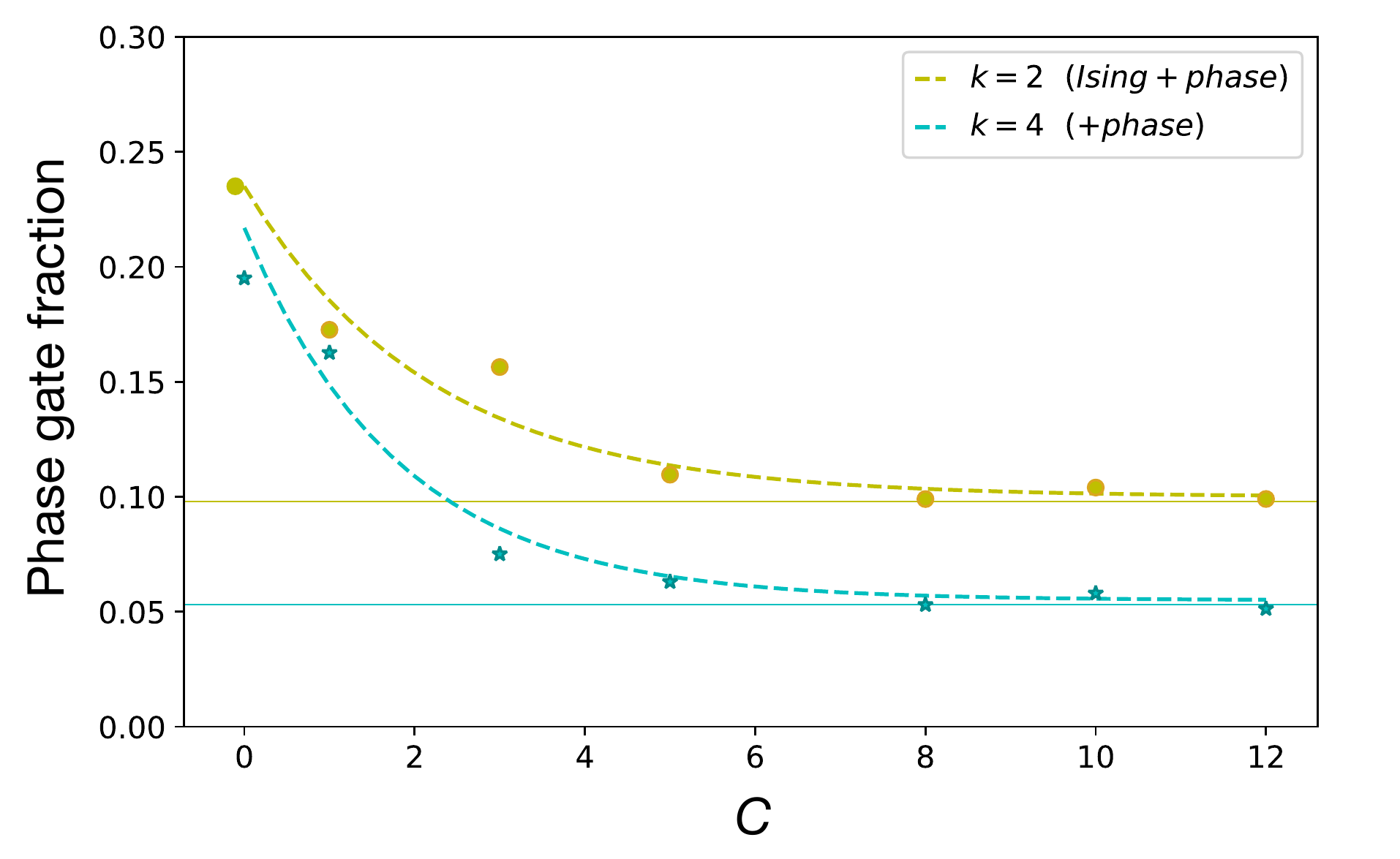}
    \caption{Ratio $N_\phi/l$ of the number of noisy phase gates $N_\phi$ and the braid word length $l=N_\phi+N_\sigma$, where $N_\sigma$ is the number of topologically protected gates, as functions of the acceptance parameter $c$ for complilation of a fixed length $l=50$, fixed precision $d = 5\times 10^{-3}$ braid word. Horizontal lines estimate the saturated fractions.}
    \label{phasefrac}
\end{figure}
Figure~\ref{phasefrac} shows the fraction of phase gates used  in a compiled braid word as a function of the control parameter $c$, for a fixed braid word precision. 

From this we infer that the optimised phase gate dependence saturates at values much lower than the expected $20\%$. Specifically, the $k=2$ results converge toward $10\%$ and the $k=4$ results toward $5\%$. To explain this observation, we consider the group structure generated by the braid generators. All one qubit models have two generators (and their inverses) which are cyclic. Specifically, for any generator $\sigma_i$ there exists a non-zero integer $q$ such that $\sigma_i^{q} = I$, where $I$ is the identity. As the generators are $8$-cycles and $12$-cycles in the $k=2$ and $k=4$ models, respectively, the order of the sub-groups generated by these elements independently are of the same order as the elements that generate them, i.e. $8$ and $12$. For simplicity, consider the subgroup $\Sigma_{\sigma_1}$ generated by\footnote{Note that we equally well could have chosen $\sigma_2$.} $\sigma_1$. We claim that the $k=4$ braid group is of greater order due to its cyclicity. To make this more explicit, we consider the coset structure in the two groups. Since every distinct element in the main group must belong to a coset of the $\sigma_1$ sub-group, the order of the sub-group divides the order of the main group, since each coset has to contain the same number of elements as the sub-group\footnote{This results is known as Lagrange's theorem \cite{roth2001history}.}. Thus, if $\Sigma_{\sigma_1}$ denotes the sub-group of the full group $\Sigma$ that is generated by $\sigma_1$, and $n$ denotes the index of this sub-group, i.e. $[\Sigma_{\sigma_1}:\Sigma]=n$, we may express the order of $\Sigma$  as 
\begin{equation}
    |\Sigma|= |\Sigma_{\sigma_1}| \cdot [\Sigma_{\sigma_1}:\Sigma] = |\Sigma_{\sigma_1}|\cdot n.
\end{equation}
Due to the cyclicity of the groups, not only is the order of the cyclic sub-groups $\Sigma_{\sigma_1}$ and $\Sigma_{\sigma_2}$ greater for $k=4$, but so is the index $n_{k=4}$ too. Hence the order of the $k=4$ braid group is greater than that of the $k=2$ braid group. 

Considering a sphere of radius $R$, each element in the group will occupy a solid angle $\Omega = \frac{4\pi}{|\Sigma|}$, assuming that the points are somewhat evenly distributed over the surface, which should be a fair assumption for dense gate set. This solid angle correspond to a spherical cap with circular boundary and an arc length that defines the ``curved" radius $r$. The solid angle on a sphere is $\Omega = \frac{A}{R^2}$, where $A$ is the area of the corresponding spherical cap. In spherical coordinates we find
\begin{equation}
    \Omega = \frac{1}{R^2} \int_{0}^{2 \pi} \int_{0}^{\theta} R^2 \sin(\theta) d\varphi d\theta = 2 \pi (1-\cos(\theta)).
\end{equation}
Setting this equal to $4\pi/|\Sigma|$ yields
$\theta = \arccos(1-2/|\Sigma|)$,
\begin{figure*}[!t]
  \centering
  \includegraphics[width=0.9\linewidth]{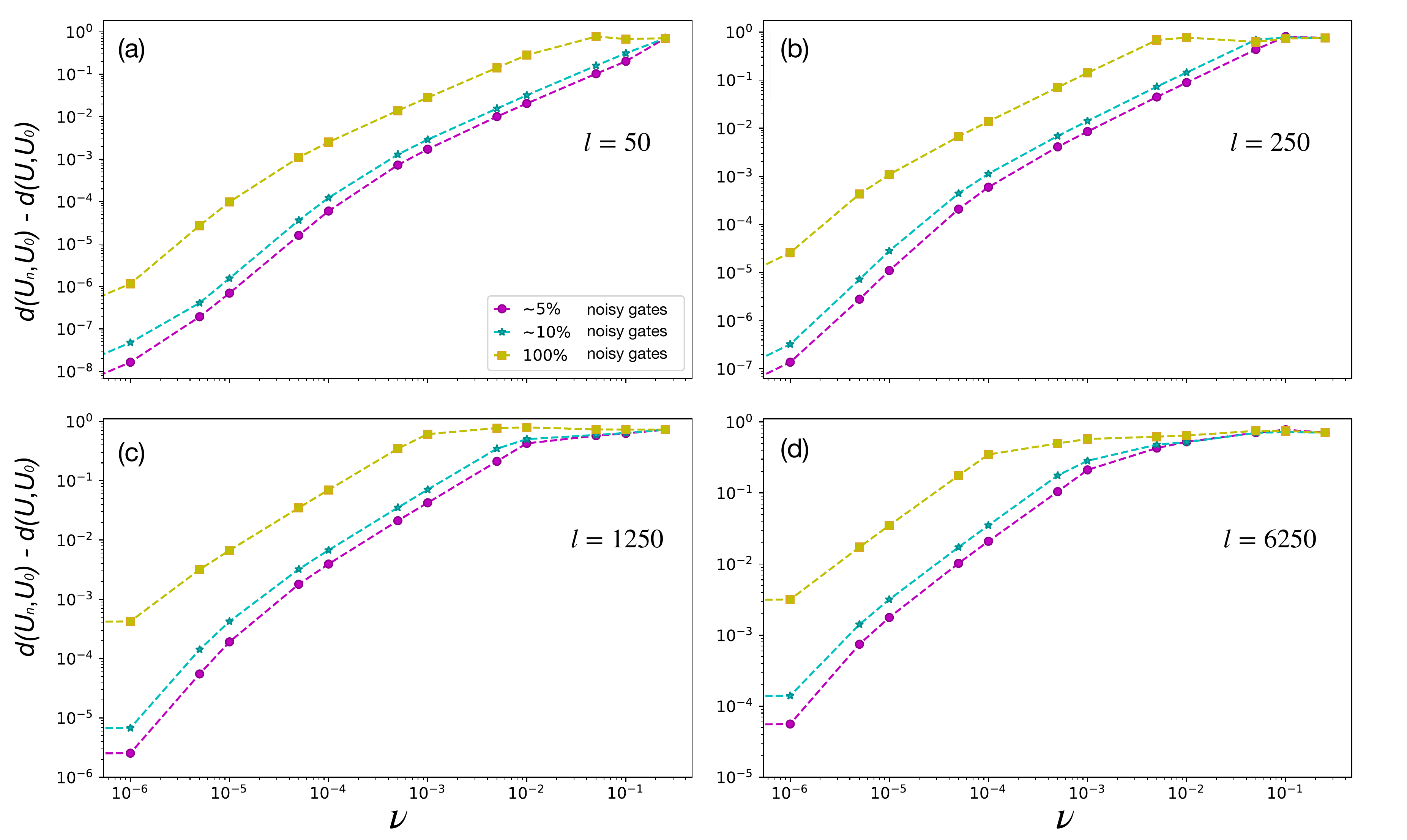}
  \caption{The braid word error purely due to the $e^{i\sqrt{2}\pi}$ phase gates deployed in the hybrid Ising anyon model as functions of the strength $\nu$ of the phase gate noise. Results (a)-(d) correspond to braid word lengths of $50, 250, 1250$, and $6250$ of Fig.~\ref{allk}.}
  \label{onlynoise}
\end{figure*}
which results in an expression for the orthodromic distance $r$
\begin{equation}
    r(|\Sigma|) = \int_{0}^{\theta} Rd\theta = R \arccos\left (1-\frac{2}{|\Sigma|}\right).
\end{equation}
This distance is the maximum distance for any point on the sphere to a group element, as each element is occupying the same area and solid angle. Note that this function goes to zero as the group order $|\Sigma|$ goes to infinity, which corresponds to the fully universal anyon models for which the target gate can be approximated to arbitrary accuracy. 

We also conclude that the average distance from any group element to any point on the sphere is shorter in the $k=4$ model than in the $k=2$ one since $|\Sigma_{k=4}|>|\Sigma_{k=2}|$. This further implies a weaker phase gate dependence for $k=4$, in agreement with the results presented in Fig.~\ref{phasefrac}. 

According to conventional wisdom, in decoherence prone quantum computation about 1000 physical qubits are required per one logical qubit to facilitate error correction \cite{heeres2017implementing,jones2018logical}. In contrast, our results suggest that the hybrid Ising anyon model would only require about 100 physical qubits for each logical one, and the $k=4$ model about 50, in order to suppress the logical error to arbitrarily low levels. 
Figure~\ref{onlynoise} demonstrates the effect of decoherence on the compiler error when noise is added to $5 \%, ~10 \%,$ or $100 \%$ of the gates. The $100 \%$ case corresponds to a $100\%$ conventional quantum computation with the Ising anyon model, for which none of the generators would possess intrinsic topological protection, or they would all be prone to strong topological decoherence at hardware level e.g. due to quasiparticle poisoning \cite{sarma2015majorana}. 
This case is included as a benchmark and reference to conventional quantum computing platforms. 

The decoherence noise is modeled as small random unitary rotations. In the $\mathfrak{su}(2)$ basis we can express the corresponding operator as 
\[U_{\rm noise}=e^{i\vec{\theta} \cdot \vec{\sigma}},
\]
where $\vec{\sigma}$ is a vector of Pauli matrices and 
 $\vec{\theta} = (\theta_1, \theta_2, \theta_3)$. We sample the  parameters $\theta_i$ from a narrow normal distribution with a zero mean, i.e. $\theta_i \in N(0,\nu)$, where $\nu$ is the standard deviation. By increasing $\nu$, greater phase gate noise fluctuations are allowed, which means that the standard deviation may be interpreted as the ``strength" of the noise. An in depth analysis of the impact of various specific kinds of noise sources is provided in \cite{conlon2019error}.

\subsection{Noise corrupted braid words in the hybrid Ising anyon model}

\begin{figure*}[!th]
  \centering
  \includegraphics[width=0.85\linewidth]{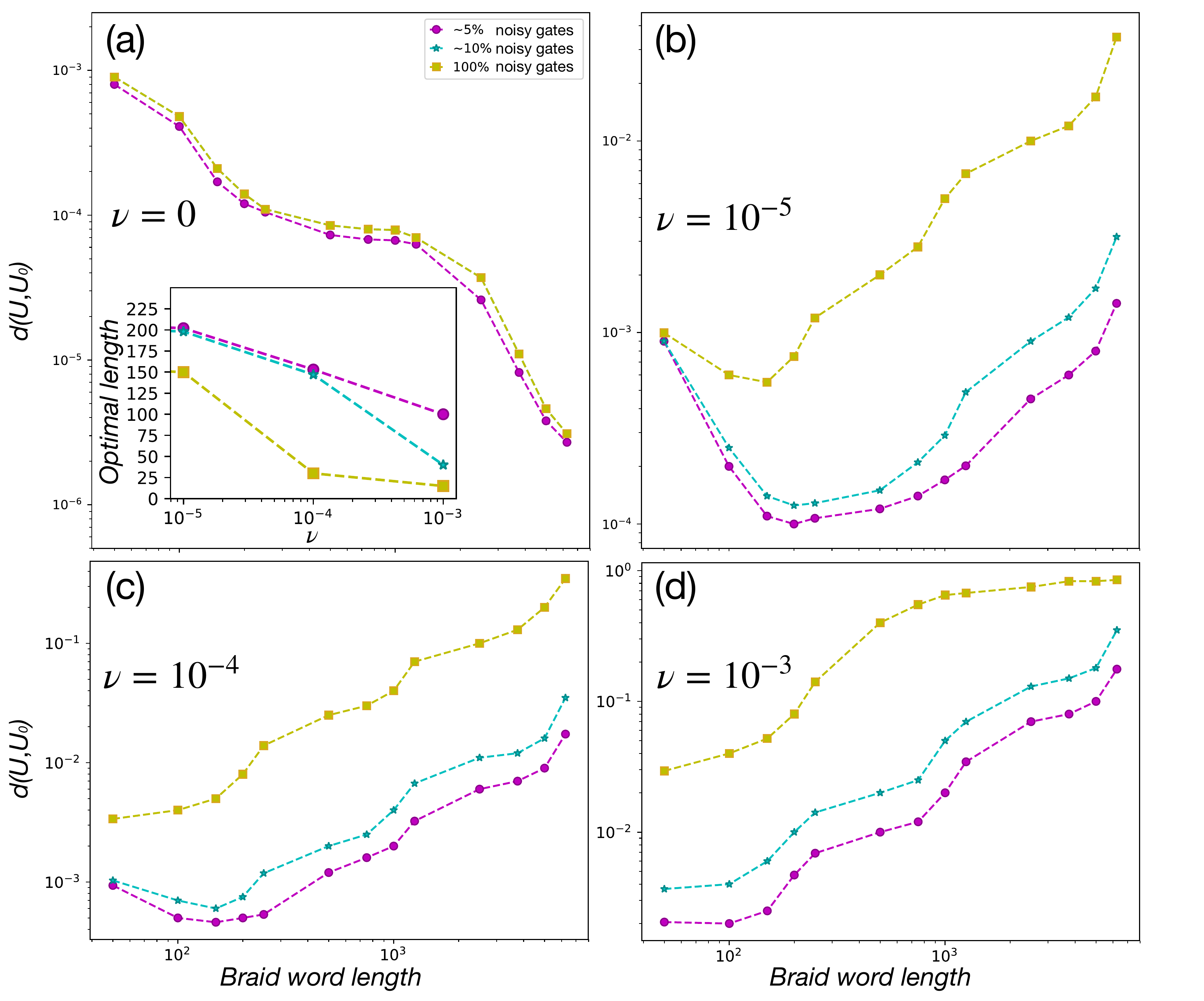}
  \caption{Absolute error including the inherent braid word approximation and the contributions from the noise in Fig.~\ref{onlynoise} as functions of the braid word length (a)-(d). The frames (a)-(d) correspond to the noise levels $\nu = 0, 10^{-5}, 10^{-4}, 10^{-3}$, respectively. The inset in (a) shows the optimal braid word length corrresponding to the minima in (a)-(d), as functions of the noise strenth $\nu$.}
  \label{errortot}
\end{figure*}

Figure~\ref{errortot} shows the results when the noise is applied to the braid words. We set $\nu = \{0, 10^{-5}, 10^{-4}, 10^{-3}\}$ 
in Fig.~\ref{errortot}(a)-(d), respectively. In the $\nu = 0$ case the noise is absent and the MCESKA should be applied to achieve the best absolute accuracy. As $\nu$ is gradually increased, error minima first appear and are continuously shifting toward shorter braid word lengths, Fig.~\ref{errortot}(b)-(c), until $\nu = 10^{-3}$ is reached in Fig.~\ref{errortot}(d), for which the error minima are obtained at the zeroth depth in the algorithm for $l=50$. Hence, for the noisy gate region the best results would be achieved with the direct Monte Carlo method. 

Summarising, in the zero noise case the full power of the Monte Carlo enhanced Solovay--Kitaev algorithm should be employed, whereas in the intermediate noise region a small number of recursions could be beneficial. For $\nu \geq 10^{-3}$ the plain Monte Carlo method should be invoked for the best results. Note that $\nu=10^{-3}$ is still quite small so it is reasonable to assume that the noise will be in this range or greater in realistic near-future applications. 

The inset in Figure~\ref{errortot}(a) shows the optimal braid word length, the location of the minima in (a)-(d), as functions of the phase gate noise $\nu$. As the noise gets stronger the optimal braid word length gets monotonically shorter. This highlights the important observation that although in a decoherence free universal TQC, such as the Fibonacci anyon model, arbitrary accuracies can in principle be achieved just by increasing the length of the braid words, in realistic applications the best attainable accuracy is always noise limited and no further benefit can be gained from deploying excessively long braid words.  

\section{Conclusions}\label{sec:conclusions}
We have developed a generic, Monte Carlo enhanced Solovay--Kitaev algorithm (MCESKA), quantum compiler to search for braid word approximations to quantum logic gates. Motivated by the potential of Majorana fermion quasiparticle zero modes as a physical realisation of the Ising anyon model, we deployed the MCESKA to assess and compare the performance and potential of the Fibonacci and the Ising anyon models for topological quantum computation. Furthermore, we expanded our analysis to include the level $k=4,5,6,8$ ${\rm SU}(2)_k$ anyon models. We found that the plain Monte Carlo quantum compiler outperformed the brute force search method in terms of efficiency and required resources while achieving comparable accuracy, and when combined with the Solovay--Kitaev algorithm (SKA), leveraged the performance by two orders of magnitude, compared to Dawson's and Nielsen's implementation of the Solovay--Kitaev algorithm in \cite{dawson2005solovay}. Although other implementations that are even more efficient than the basic SKA exist, they are non-generic, model specific algorithms \cite{kliuchnikov2013synthesis, paetznick2013repeat, selinger2012efficient, bocharov2015efficient, ross2014optimal, kliuchnikov2015practical, paetznick2013repeat}. One of the major benefits of the MCESKA is that it retains full model independent generality while providing efficient means to perform quantum gate compilation. Moreover, the MCESKA is a versatile algorithm that can easily accommodate arbitrary constraints on the braid words. For instance, an inclusion of a phase gate acceptance criterion allowed us to suppress the noisy phase gate count significantly. It is also directly applicable to multi-qubit systems with arbitrary number of braid generators.

Presently, physical systems hosting Majorana fermion zero modes appear to be among the most promising platforms for TQC. Since the resulting Ising anyon model cannot achieve universal quantum computation by braiding alone, it means that such systems will have to also deal with conventional forms of decoherence. Our results suggest that the hybrid anyon models' susceptibility to conventional types of decoherence due to environmental noise can be reduced by efficient gate compilation that minimizes the use of the auxliary, noisy phase gates. The seemingly ineffective raw Ising anyon model can be made computationally universal by adding an irrational phase gate to its generator set. This extra gate drastically enhances the quality of the generator set and the results presented in Fig.~\ref{fibvsising} surprisingly indicate that it could actually perform slightly better than the inherently universal Fibonacci model that enjoys full benefits of topological protection. 

However, when phase gate noise is introduced to the anyon models, we found that even in the case of rather weak noise, the use of long braid words results in larger compilation errors than the shorter ones because of the necessary use of a larger number of noisy phase gates. The downside of this is that the theoretical accuracy of the hybrid topological quantum computation is limited by the conventional kinds of noise, while the upside is that relatively short braid words of only tens or hundreds of braidings can be used, which is good news from the future hardware implementation point of view.

Most importantly, where as a fully conventional quantum computer is estimated to require of the order of thousand physical qubits per each logical qubit due to having to dispense vast resources on error correction protocols, a hybrid topological quantum computer based on Ising anyons would only need an order of magnitude fewer in order to achieve comparable computational accuracy. This demonstrates that even the hybrid topological quantum computer models retain a clear topological advantage over the conventional kinds of quantum computers.

In conclusion, that a hybrid Ising anyon model may, under similar circumstances, perform significantly better than a conventional decoherence prone quantum computer, seems promising for the quest of realising Majorana fermion based topological quantum computers.

\begin{acknowledgements}
We are grateful to Chris Vale for useful discussions. This work was performed on the OzSTAR national facility at Swinburne University of Technology. The OzSTAR program receives funding in part from the Astronomy National Collaborative Research Infrastructure Strategy (NCRIS) allocation provided by the Australian Government.
This research was funded by the Australian Government through the Australian Research Council (ARC) Future Fellowship project FT180100020.
\end{acknowledgements}

\bibliography{apssamp}

\end{document}